%% file: group_LF.tex
\documentclass[fleqn,usenatbib,usedcolumn]{mnras}
\usepackage[british]{babel}             
\usepackage{newtxtext}                  
\usepackage{newtxmath}                   
\let\la=\lesssim 
\let\ga=\gtrsim
\usepackage[T1]{fontenc}                
\usepackage{bm}
\usepackage[pdftex]{graphicx}
\usepackage{mathtools}
\usepackage{times}
\usepackage{url}

\newcommand{\gtc}{\mbox{G$^3$C}}
\newcommand{\hMpc}{\mbox{\,$h^{-1}$\,Mpc}}

\newcommand{\mass}{\mbox{${\cal M}$}}

\newcommand{\volunit}{\mbox{\,$h^{-3}$\,Mpc$^{3}$}}

\newcommand{\Sersic}{S\'{e}rsic}

\newcommand{\Vmax}{\mbox{$V_{\rm max}$}}
\newcommand{\Vdcmax}{\mbox{$V^{\rm dc}_{\rm max}$}}

\newcommand{\lgal}{\mbox{\sc L-Galaxies}}
\newcommand{\tng}{\mbox{IllustrisTNG}}

\title[GAMA: group galaxy LF and SMF]
{Galaxy and Mass Assembly: luminosity and stellar mass functions
  in GAMA groups}

\input authors.tex

\date{Accepted XXX. Received YYY; in original form ZZZ}

\pubyear{2020}

\begin{document}
\label{firstpage}
\pagerange{\pageref{firstpage}--\pageref{lastpage}}
\maketitle

\begin{abstract}
  How do galaxy properties (such as stellar mass, luminosity,
  star formation rate, and morphology) and their evolution
  depend on the mass of their host dark matter halo?
  Using the Galaxy and Mass Assembly (GAMA) group catalogue,
  we address this question by exploring the dependence on host halo mass
  of the luminosity function (LF)
  and stellar mass function (SMF) for grouped galaxies subdivided by colour,
  morphology and central/satellite.
  We find that spheroidal galaxies in particular dominate the bright and
  massive ends of the LF and SMF, respectively.
  More massive haloes host more massive and more luminous central galaxies.
  The satellite LF and SMF respectively show a systematic
  brightening of characteristic magnitude, and increase in characteristic mass,
  with increasing halo mass.
  In contrast to some previous results,
  the faint-end and low-mass slopes show little systematic dependence
  on halo mass.
  Semi-analytic models and simulations show similar or enhanced dependence
  of central mass and luminosity on halo mass.
  Faint and low-mass simulated satellite galaxies are remarkably independent
  of halo mass, but the most massive satellites are more common in more
  massive groups.
  In the first investigation of low-redshift LF and SMF evolution
  in group environments,
  we find that the red/blue ratio of galaxies in groups has increased since
  redshift $z \approx 0.3$ relative to the field population.
  This observation strongly suggests that quenching of star formation
  in galaxies as they are accreted into galaxy groups is a significant and
  ongoing process.
\end{abstract}


\section{Introduction}

In the hierarchical model of galaxy formation, haloes of dark matter (DM)
grow by gravitational attraction and merging to form larger haloes
\citep[e.g.][]{Press1974,White1978}.
These haloes also attract baryons,
a small fraction of which will condense into stars and thence form galaxies.
How do galaxy properties, such as stellar mass, luminosity,
star-formation rate, and morphology, depend on host halo mass and
evolutionary history?
The connection between galaxies and their host DM haloes is an active area
of astrophysical research \citep[see][for a recent review]{Wechsler2018}.
One approach to studying this connection is to identify and weigh individual
haloes using galaxies as tracers.
Galaxy group catalogues provide a way to estimate the total mass
of individual haloes down to $\sim 10^{12} \mass_\odot$
via the (assumed virialized) galaxy motions within them
\citep{Eke2006,Robotham2011},
or by weak-lensing calibrated scaling relations \citep{Han2015,Viola2015}.

The galaxy luminosity function (LF) and stellar mass function (SMF) are
fundamental observables,
giving a description of the population of galaxies in
different environments, and contain valuable information about the
physical processes that feature prominently in galaxy formation and
evolution.
The LF and SMF and their evolution provide important constraints on
theories and models of galaxy formation and evolution
\citep[e.g.][]{Benson2003,Gonzalez-Perez2014,Lacey2016,Lagos2018}.

In the last few years, many authors have investigated the effect of
environment on the LF, focusing on the dependence of the LF on the
density contrast within spheres of different radii
\citep[e.g.][]{Croton2005,Hoyle2005,Xia2006,Park2007,Phleps2007,McNaught-Roberts2014}.
These works agree that the LF varies significantly with environment,
with characteristic magnitude brightening systematically with
increasing local density.
What is less clear is any systematic dependence of the faint-end slope
with density, with some authors \citep[e.g.][]{Xia2006}
claiming a steepening slope
(i.e. more dwarf galaxies) in higher-density environments,
while others \citep[e.g.][]{Croton2005,Hoyle2005,McNaught-Roberts2014}
see little correlation.
The SMF as a function of projected density has been presented by
\citet{Peng2010}, who find that the low-mass SMF of red galaxies is slightly
steeper in the highest density quartile, while the low-mass slopes for blue
galaxies are indistinguishable.
While \citet[fig.~14]{Mortlock2015} find a steeper SMF slope in
high-density environments at redshifts $z \ga 0.5$,
they find the opposite in their low-redshift bin.
Earlier, \citet{Baldry2006} found that characteristic mass increases
with projected density.

Large spectroscopic surveys of galaxies, such as the Sloan Digital Sky Survey
(SDSS; \citealt{York2000}) and the 2dF Galaxy Redshift Survey (2dFGRS;
\citealt{Colless2001}) provide the potential for group-finding
based on the redshift-space distribution of galaxies.
Many authors have taken advantage of these surveys to construct galaxy group
catalogues to explore multiple aspects of these systems,
\citep[e.g.][]{Merchan2002,Merchan2005,Eke2004,Yang2005,Yang2007,Berlind2006,Weinmann2006,Munoz-Cuartas2012}.
In particular, the dependence of the galaxy LF on group environment
has been investigated by, e.g.
\citet{Eke2004a,Robotham2006,Robotham2010,Zandivarez2006,Zandivarez2011,Guo2014c}.
These works mainly explored the variation of the \citet{Schechter1976}
function parameters,
the characteristic magnitude $M^*$ and the faint-end slope $\alpha$,
for different galaxy populations, as a function of the
galaxy group virial mass, multiplicity, velocity dispersion,
etc. Their results showed clear variations of $M^*$ and $\alpha$
with the different group properties.
\citet{Robotham2010} found clear
trends for steepening faint-end slope $\alpha$ as group mass and/or
multiplicity increase
for early-type galaxies, while a much suppressed relation was observed
for the late-type population.
\citet{Zandivarez2011} found similar results.

Rather than measuring the number density of galaxies per unit volume,
one can instead measure the average number of galaxies per host group
\citep[e.g.][]{Yang2003}.
The conditional luminosity function (CLF), $\phi_C(L | \mass_h)$,
describes the average number of galaxies as a function of luminosity $L$
in groups of mass $\mass_h$, i.e. average number \emph{per group} rather than
per unit volume, and can be considered an extension of the
halo occupation distribution (HOD) model \citep[e.g.][]{Berlind2002,Brown2008}.
Similarly, the conditional stellar mass function (CSMF),
$\phi_C(\mass_* | \mass_h)$,
describes the average number of galaxies per group as a function of their
stellar mass $\mass_*$.
Using the SDSS DR4 catalogue,
\citet{Yang2008,Yang2009} found that the characteristic
luminosity gets brighter, the characteristic mass increases,
and the faint- and low-mass slopes of the CLF and CSMF get steeper,
as halo mass increases.
There is a danger, however, in characterising LF dependence on environment
purely in terms of Schechter function parameters.
The Schechter parameters $(\alpha, M^*)$ are strongly correlated,
and also very sensitive to the limiting magnitude used in the fit
\citep[appendix C]{Croton2005}.
Thus the Schechter function parametrization should only be used if
(i) the fit is performed over a consistent magnitude range,
and (ii) the functional fit is a good one (as confirmed by a $\chi^2$-test
or likelihood ratio comparison with a non-parametric estimate).

The Galaxy and Mass Assembly
(GAMA; \citealt{Driver2009,Driver2011,Liske2015}) survey provides an
opportunity to reassess the galaxy LF and SMF
dependence on host group properties.
Although of smaller area than SDSS, GAMA provides spectroscopic redshifts
two magnitudes fainter than SDSS, and, even more importantly
for group studies, is highly complete, even in high-density group environments.
The dependence of the galaxy LF on local environment,
as defined by galaxy counts in $8 \hMpc$ spheres,
has previously been presented for GAMA data by
\citet{McNaught-Roberts2014},
who found that denser environments contain redder and brighter galaxies
than low-density environments.
\citet{Alpaslan2015} carried out a wide-ranging exploration of the effects
of environment, including host group mass, on galaxy properties,
finding that the characteristic stellar mass increases with group mass.
\citet{Barsanti2018} and \citet{Wang2018} have recently investigated
the impact of GAMA group environment on star formation.
\citet{Barsanti2018} find that the fraction of star-forming galaxies
is higher in group outskirts where galaxies have recently been accreted,
and lower in the central, virialized regions.
\citet{Wang2018} find that, overall,
star formation rate is suppressed in group environments relative to the field.

In this paper, we present galaxy LFs and SMFs as a function of host group mass,
subdivided by galaxy colour, morphology, and by redshift.
In Section~\ref{sec:data} we describe the GAMA Galaxy Group Catalogue (\gtc)
and associated galaxy samples, as well as comparison mock catalogues and
simulations.
Section~\ref{sec:method} describes the methods used to estimate
the LFs and SMFs in bins of halo mass and redshift.
Section~\ref{sec:results} shows our results and we conclude in
Section~\ref{sec:concs}.
In Appendix~\ref{sec:field_comp} we compare field LFs and SMFs between
GAMA and mock and simulated samples.
Appendix~\ref{sec:fofvhalo} investigates the effects of group-finding and
halo mass estimation by comparing LFs using true mock groups and
masses with those based on estimated quantities.
We test our estimators on simulated data in Appendix~\ref{sec:simtests},
showing that the $1/\Vmax$-weighted LF provides unbiased estimates,
whereas the per-group CLF is biased in low-mass groups unless stringent
redshift cuts are imposed.

For this work, we assume cosmological parameters of
$\Omega_M=0.3$, $\Omega_{\Lambda}=0.7$ with a Hubble constant of
$H_0=100 h$ km s$^{-1}$ Mpc$^{-1}$.
Group (halo) masses have been calibrated by weak lensing measurements,
and are represented on a logarithmic scale by
$\lg \mass_h \equiv \log_{10} (\mass_h/\mass_\odot h^{-1})$.
Stellar masses in simulations, whose natural units are $\mass_\odot h^{-1}$,
are scaled by the relevant value of $h$ to be consistent with
stellar masses for observed galaxies, so that both are represented by
$\lg \mass_* \equiv \log_{10} (\mass_*/\mass_\odot h^{-2})$.

\section{GAMA data, mocks and simulations} \label{sec:data}

The GAMA project is a multi-wavelength
spectroscopic galaxy survey based on an input catalogue described by
\cite{Baldry2010}.
In this paper, we make use of the GAMA-II \citep{Liske2015} equatorial fields,
each of 12 $\times$ 5 degrees centred at 09h, 12h
and 14h30m RA, called G09, G12 and G15 respectively. The GAMA-II Petrosian
magnitude limit is $r < 19.8$ mag for all three fields. This survey is
complete in all regions with a completeness greater than 96\% for all
galaxies with up to 5 neighbours within 40 arcsec
\citep[see][for a detailed description]{Liske2015}.
We first discuss the GAMA mock catalogues,
as these are used to justify our choice of group mass estimator.

\subsection{Mock catalogues and group mass estimates}

The GAMA mock catalogues have been
designed to match GAMA-I survey data as closely as possible
(updates to reflect the extended area of GAMA-II are currently in progress).
These were constructed from the Millennium dark matter simulation
\citep{Springel2005} and
populated with galaxies using the {\sc galform} \citep{Bower2006}
semi-analytic galaxy formation recipe.
They are the same mocks used to tune and test the GAMA group-finding algorithm
in \citet[][hereafter R11]{Robotham2011};
readers are referred to that publication for further details
of the mock GAMA group catalogues.

We compare the LFs of the GAMA mocks with GAMA data in
Appendix~\ref{sec:field_comp}, finding that the characteristic magnitude
of the mock galaxies is about 0.5 mag fainter than for GAMA galaxies.
When comparing GAMA and mock grouped LFs,
one should therefore focus on the trends with halo mass for each,
rather than compare LF parameters.

Two mock group catalogues are available.
The first, {\tt G3CMockHaloGroupv06},
hereafter referred to as \emph{halo mocks},
contains the positions and masses $\mass_{\rm halo}$
of the intrinsic haloes in the dark matter simulations.
The second, {\tt G3CMockFoFGroupv06}, referred to as \emph{FoF mocks},
has groups identified using the same friends-of-friends (FoF) algorithm,
and masses estimated in the same way as for the GAMA data.

We compare two methods for estimating group masses.
The first derives a dynamical mass $\mass_{\rm dyn}$ via the virial theorem
from galaxy dynamics within each group (column {\tt MassA} in the
relevant group catalogue).
The second derives a luminosity-based mass $\mass_{\rm lum}$
from group $r$-band luminosity
(column {\tt LumB}) using the weak-lensing calibrated
scaling relation of \citet[][eqn.~37]{Viola2015}.
{\tt LumB} provides the total $r$-band luminosity down to
$M_r - 5 \log_{10} h = -14$ mag
in solar luminosities, multiplied by a constant calibration factor of $B=1.04$
(see R11 section 4.4 for details)\footnote{The GAMA and mock group catalogues
  include an alternative group luminosity
estimate, {\tt LumBfunc}, in which the calibration factor $B$ is a function
of redshift and group multiplicity.
However, the GAMA and mock groups show 
significantly discrepant distributions of {\tt LumBfunc},
with mock galaxies being on average about 1.6 times more luminous
than GAMA galaxies.
We also note (Margot Brouwer, private communication)
that the \citet{Viola2015} scaling relations use {\tt LumB}
and not {\tt LumBfunc}.}.

\begin{figure} 
\includegraphics[width=\linewidth]{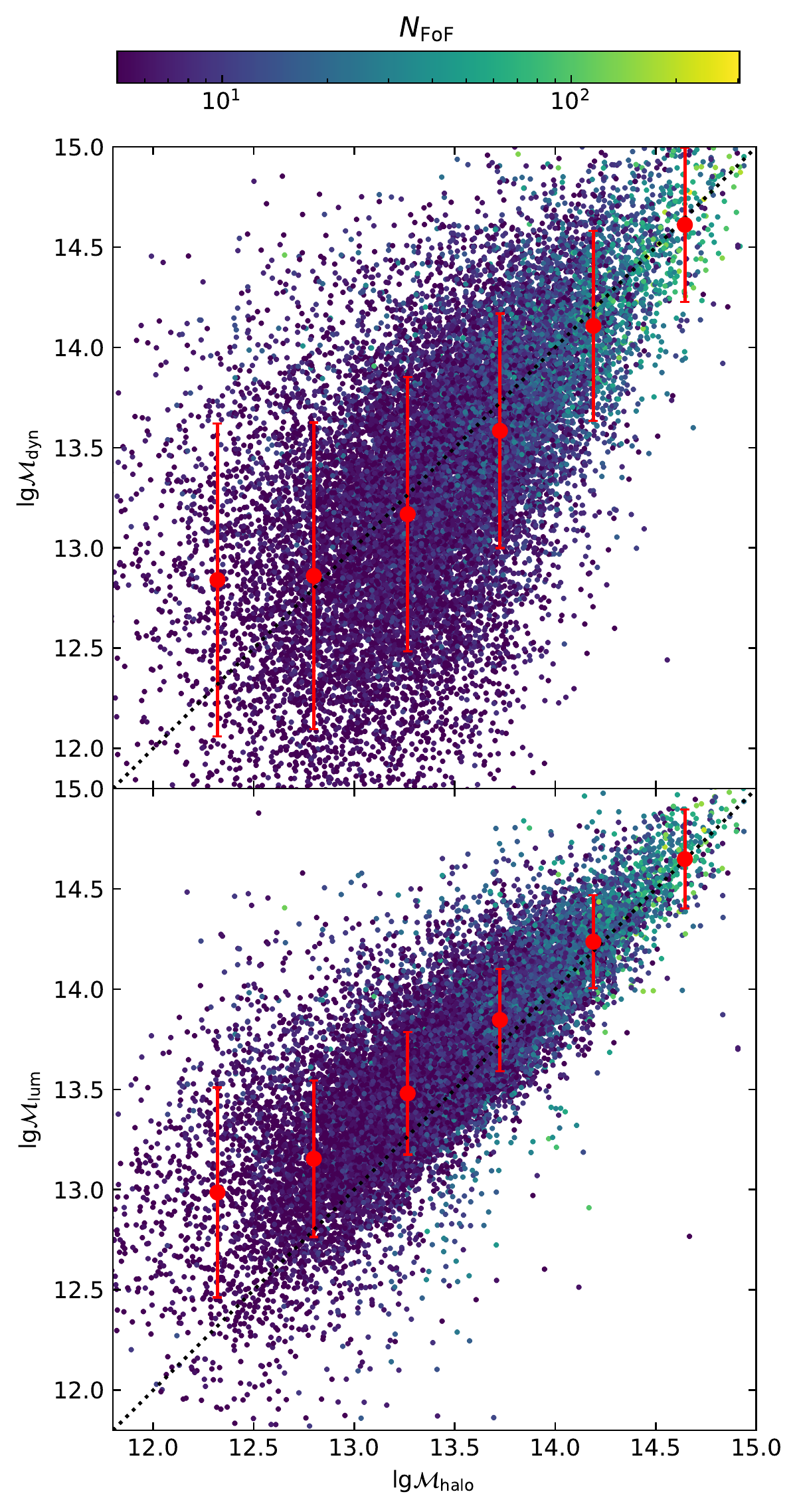} 
\caption{Comparison of luminosity-based ($\lg {\cal M}_{\rm lum}$, lower panel),
  and dynamical ($\lg {\cal M}_{\rm dyn}$, upper panel), estimates of
  mock group mass, against true mock halo mass, $\lg {\cal M}_{\rm halo}$,
  colour coded by group membership.
  See text for details of these mass estimates.
  The red error-bars show mean and standard deviation of estimated halo mass
  in 0.5 mag bins of $\lg {\cal M}_{\rm halo}$.
}
\label{fig:mass_comp}
\end{figure}

In order to check the reliability of these mass estimates,
we match groups in the mock halo catalogue with those in the mock FoF catalogue
on the basis of sharing the same iterative centre
(see R11 section 4.2 for the definition of this).
As for the real GAMA groups, we select only mock FoF groups with
five or more members,
as these richer groups are found to be the most reliable (R11).
We also exclude groups for which less than 90\% of the group is estimated
to lie within the survey boundaries, i.e. we require {\tt  GroupEdge} $> 0.9$.
We can then compare the luminosity- and dynamically-based mass estimates
from the FoF catalogue with the true halo masses from the halo catalogue.
In Fig.~\ref{fig:mass_comp}, we see that the luminosity-based masses
(lower panel) show a better correlation with halo mass than do the dynamical
mass estimates (top panel), in agreement with the results of \citet{Han2015}.
We therefore use only the luminosity-based mass estimates in this paper.
We note that both estimators are biased high at low halo masses,
with a more pronounced bias for $\mass_{\rm lum}$ due to its smaller scatter.
This suggests that the FoF group finder is tending to include spurious
members in lower-mass groups, a perhaps not unexpected result
given that the FoF linking length is independent
of halo mass (cf. the halo-based group finder used for the \citealt{Yang2007}
group catalogue, in which linking length scales with halo mass).
When interpreting the dependence of galaxy luminosity on halo mass,
one should also  bear in mind that estimated halo mass is based on
integrated galaxy luminosity.
This circular logic is also true of previous work
\citep[e.g.][]{Yang2008,Yang2009}.

Uncertainties on mock LF estimates are determined from the scatter between
nine independent realisations of the GAMA-I survey volume
(each realisation comprising three $12 \times 4$ deg regions;
20 per cent smaller than the GAMA-II equatorial fields).
Mock galaxies are taken from {\tt G3CMockGalv06}.
Absolute magnitudes are $K$-corrected (to redshift zero) with
universal $K$- and $e$-corrections as specified in Sec.~2.2 of R11.
These GAMA mocks do not provide colour or morphological information
for the galaxies,
and so we present only `total' mock LFs, without subdivision by colour or
\Sersic\ index.
Neither do these mock catalogues include stellar mass estimates,
so we are unable to compare SMFs.
Instead, we compare SMFs with the \lgal\ semi-analytic model,
and two hydrodynamical simulations, described below.

\subsection{GAMA group data}

The GAMA Galaxy Group Catalogue (\gtc v9) was generated using the
GAMA-II spectroscopic survey and applying a friends-of-friends (FoF)
grouping algorithm; the first version of this catalogue (\gtc v1) is
presented by R11 using the GAMA-I survey.
The \gtc v9 (hereafter abbreviated to \gtc) 
catalogue contains a total of 23,654 groups (comprising 2 or more members)
containing a total of 75,029 galaxies;
$\sim 40\%$ of GAMA galaxies are assigned to groups.
As for the mocks, we utilise only groups which have five or more
member galaxies and {\tt GroupEdge} $> 0.9$.
This leaves us with a sample of 24,832 galaxies in 2,718 groups.

\begin{figure}
\includegraphics[width=\linewidth]{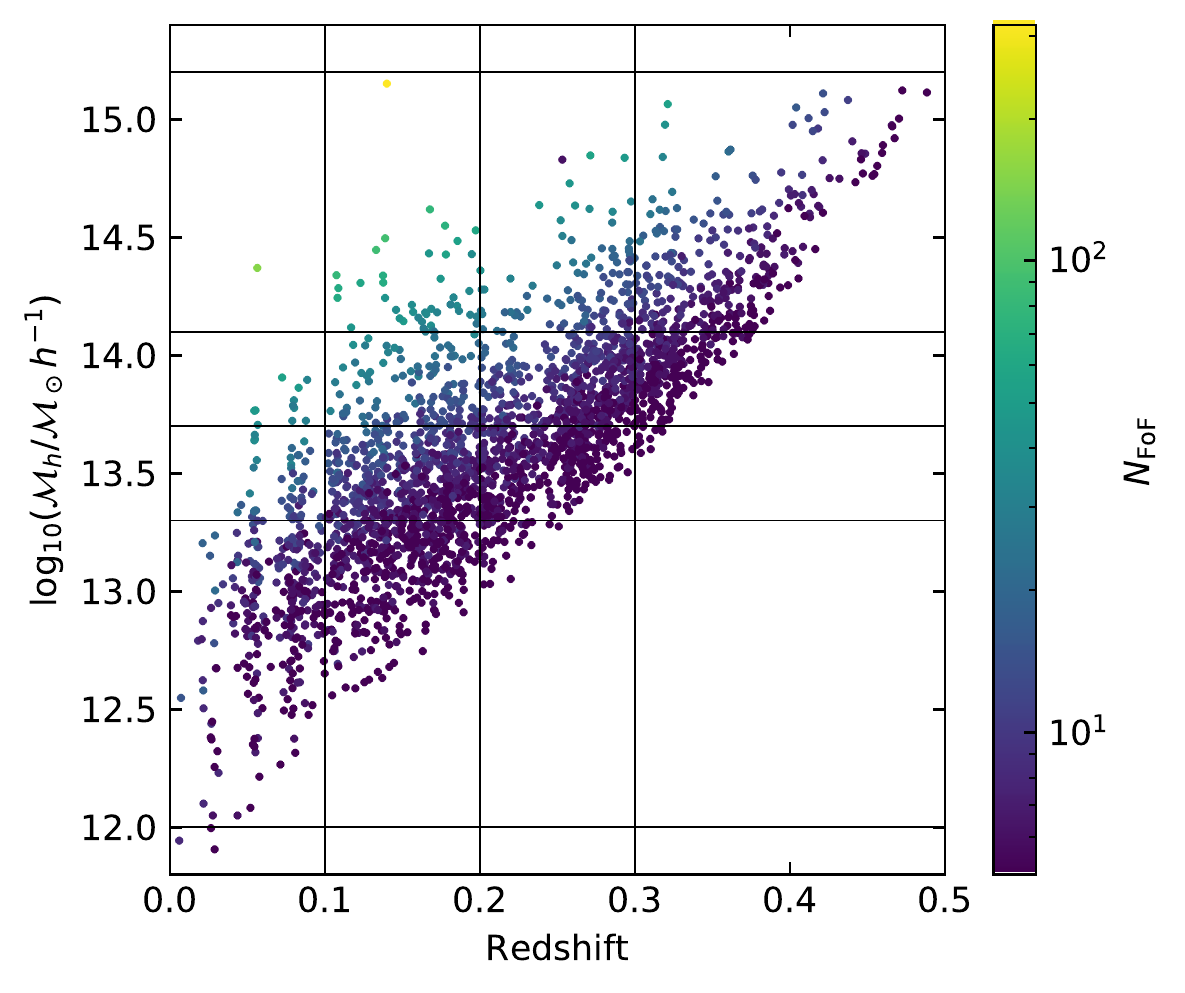} 
\caption{Mass--redshift distribution for GAMA groups that satisfy
  our selection criteria.
  Colour-coding indicates the number of group members on a logarithmic scale.
  The horizontal lines delineate the halo mass bins used in this analysis
  and the  vertical lines show the redshift bins used when investigating
  LF evolution.}
\label{fig:mass_z}
\end{figure}

Masses are estimated from group luminosities {\tt LumB}
via the \citet{Viola2015} scaling relation,
as discussed in the previous subsection.
The mass--redshift distribution of our selected GAMA groups is shown in
Fig.~\ref{fig:mass_z}.
There is a clear selection bias against finding low-mass groups at
high redshift, demonstrating a strong correlation between group mass
and the $r$-band luminosity of its fifth-brightest member.
It is also, unsurprisingly, apparent that higher-mass groups tend to have more
observed galaxy members.
Groups at higher redshift for fixed mass tend to have fewer members, simply
due to the $r < 19.8$ mag flux limit of the GAMA-II survey.


\input group_mass.tex

We sub-divide the groups into four mass bins as defined in
Table~\ref{tab:group_mass_def},
chosen to provide roughly comparable numbers of galaxies.
Comparing the halo and FoF mock groups, it is clear that the FoF algorithm
is systematically overestimating the numbers of groups in all mass bins.
It seems likely that the higher numbers of FoF cf. halo groups
is due to the FoF algorithm aggregating lower-mass haloes,
which individually would not satisfy our selection criteria, into one system.
Altogether, the FoF mock catalogue contains about 20 per cent more groups
with five or more members than does the halo mock catalogue.
The numbers of GAMA group in each bin lie somewhere between the halo and
FoF mocks, bearing in mind the 20 per cent smaller volume of the mocks.

In Appendix~\ref{sec:fofvhalo}, we investigate the effects of FoF group finding
and luminosity-based mass estimation, by comparing
LFs obtained from halo and FoF mock catalogues.
We find that while the halo and FoF non-parametric LFs show qualitatively
similar behaviour, 
they are formally inconsistent in all but the lowest mass bin, 
and with Schechter parameters that disagree by about 1--3$\sigma$.
It is likely that our GAMA results will suffer from similar biases.

\subsection{Galaxy data}

\subsubsection{Central versus satellite}

Galaxies assigned to each group are ranked according to distance from the
iterative centre of the group (R11, section 4.2.1).
We define the first-ranked galaxy in each group as the central galaxy
(95 per cent of the time this corresponds to the brightest galaxy),
and all other galaxies as satellites, so that each group has one
central galaxy and four or more satellites.
Note that the GAMA group catalogue is constructed using a friend-of-friend (FoF)
algorithm, whereas the SDSS group catalogue of \citet{Yang2007} is constructed
using a halo-based method.
As discussed by \citet{Robotham2010}, the latter algorithm results in
groups typically containing smaller numbers of galaxies,
including groups that comprise a single galaxy,
and so our results for central and satellite galaxies are not directly
comparable with those of \citet{Yang2008,Yang2009}.
One could choose to treat ungrouped galaxies in the \gtc\ as isolated centrals,
but their host halo properties would be extremely uncertain.

\subsubsection{$r$-band luminosities}

Our $r$-band LFs are derived from SDSS DR7 Petrosian magnitudes,
corrected for Galactic extinction using the dust maps of \cite{Schlegel1998}.
Since galaxies are observed at different redshifts, a correction to
the intrinsic luminosity has to be applied according to the rest frame
of the galaxy. All galaxies in this analysis have been corrected by
the so-called $K$-correction \citep{Humason1956} using the 
{\sc kcorrect v4\_2} code (\citealt{Blanton2003,Blanton2007})
employing the SExtractor \citep{Bertin1996} AUTO magnitudes reported
in ApMatchedCatv06 \citep{Driver2016}.
These $K$-corrections were obtained from the GAMA data management unit (DMU)
{\tt kCorrectionsv05} \citep{Loveday2015}.
In order to be compatible with results from the GAMA mocks and
hydrodynamical simulations, we $K$-correct to redshift zero\footnote{Although
  simulation snapshots are at higher redshifts,
  the photometric bands are rest-frame.}.
Absolute magnitudes in this band are indicated by $^{0.0}M_r$.

When not subdividing into redshift bins,
we apply a luminosity evolution correction of $+Q_e z$ mag, where $Q_e = 1.0$.
In principle, one might expect evolution to be environment-dependent,
but due to degeneracies when simultaneously fitting for luminosity evolution,
density evolution,
and large-scale structure density variations (see \citealt{Loveday2015}),
we assume global evolution corrections.
See Section~\ref{sec:vmax_est} for more details on these evolution corrections.

\subsubsection{Stellar masses}

Galaxy stellar masses are obtained from the GAMA DMU
{\tt StellarMassesLambdarv20} \citep{Taylor2011}.
The stellar masses given in this table are based on {\sc lambdar}
matched aperture photometry \citep{Wright2016}.
We apply a correction for aperture to total flux using the {\tt fluxscale}
parameter, which gives the ratio of total (\Sersic) to {\sc lambdar} flux.
We use those 96 per cent of galaxies with a physically reasonable value of the
{\tt fluxscale} parameter, that is in the range 0.8--10.
See \citet{Wright2017} for a comparison of these stellar mass estimates,
based on optical to near-IR photometry,
with alternative estimates made using {\sc magphys}
\citep{DaCunha2008,DaCunha2011},
as well as a comprehensive discussion of possible systematic errors
affecting stellar mass estimates.

\subsubsection{Colour}

\begin{figure} 
\includegraphics[width=\linewidth]{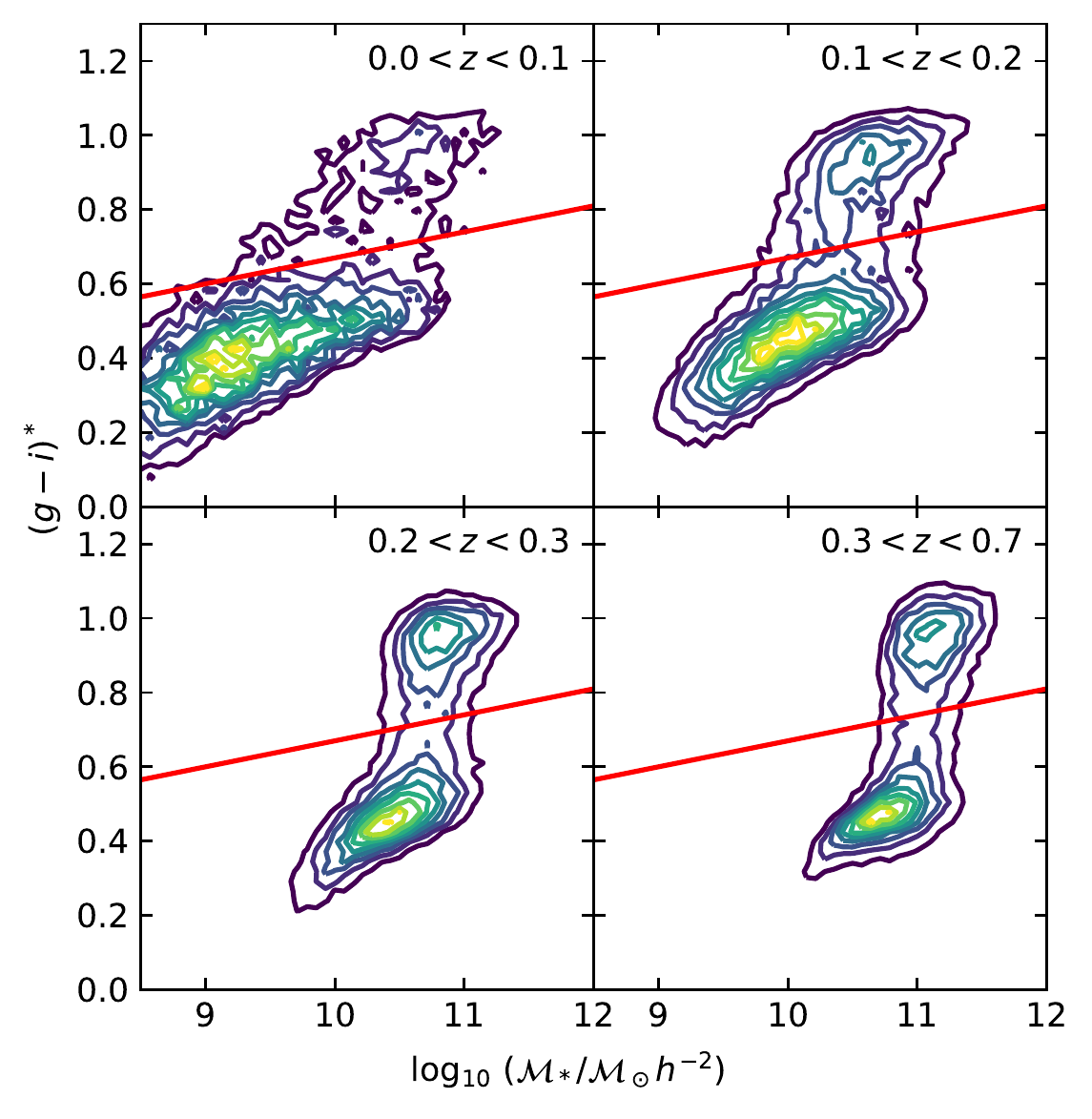} 
\caption{$(g - i)^*$ intrinsic stellar colour versus log stellar mass in four
  redshift slices as labelled.
  Contours are linearly spaced in
  density.  The red line shows our blue/red division given by equation
  (\ref{eqn:colour_cut}).}
\label{fig:colour_mass}
\end{figure}

\gtc\ member galaxies are separated into red and blue populations 
using restframe and dust-corrected $(g - i)^*$ intrinsic stellar colours
from the {\tt StellarMassesLambdarv20} DMU \citep{Taylor2011}.
In Fig.~\ref{fig:colour_mass}
we plot $(g - i)^*$ colour versus log stellar mass in four redshift slices.
The red line is a linear dividing line, fit by eye, which roughly follows
the division between 'R' and 'B' galaxies in Fig.~11 of \citet{Taylor2015},
and is given by
\begin{equation}  \label{eqn:colour_cut}
  (g - i)^* = 0.07 \log_{10} (\mass_*/\mass_\odot h^{-2}) - 0.03.
\end{equation}
Fig.~\ref{fig:colour_mass} demonstrates that this cut is applicable over the
full redshift range of the GAMA-II survey, and has the advantage that it is
corrected for internal dust-reddening.
We note that with this definition, there are very few red galaxies at
low redshift, $z < 0.1$.
\citet{Taylor2015} argue that a probablistic assignment of galaxies to
'R' and 'B' populations is preferable to a hard (and somewhat arbitrary)
red/blue cut.
However, for our purposes, dividing the galaxy population
into star-forming and quiescent using a hard cut, is quite adequate,
and certainly a lot simpler than applying the \citet{Taylor2015} 40-parameter
probabilistic model (which has been tuned for nearby $z < 0.12$ galaxies).

\subsubsection{Morphology}

The morphology of galaxies is fundamental to understanding their
behaviour at different evolutionary epochs.
We are therefore
interested in comparing spheroidal and disky galaxy shapes with colour.
Generally, red colour is associated with galaxies
containing a low fraction of dust and low star formation, i.e. early
type or spheroidals, while the blue population is usually associated
with star forming galaxies or late types, mainly spirals.

\begin{figure} 
\includegraphics[width=\linewidth]{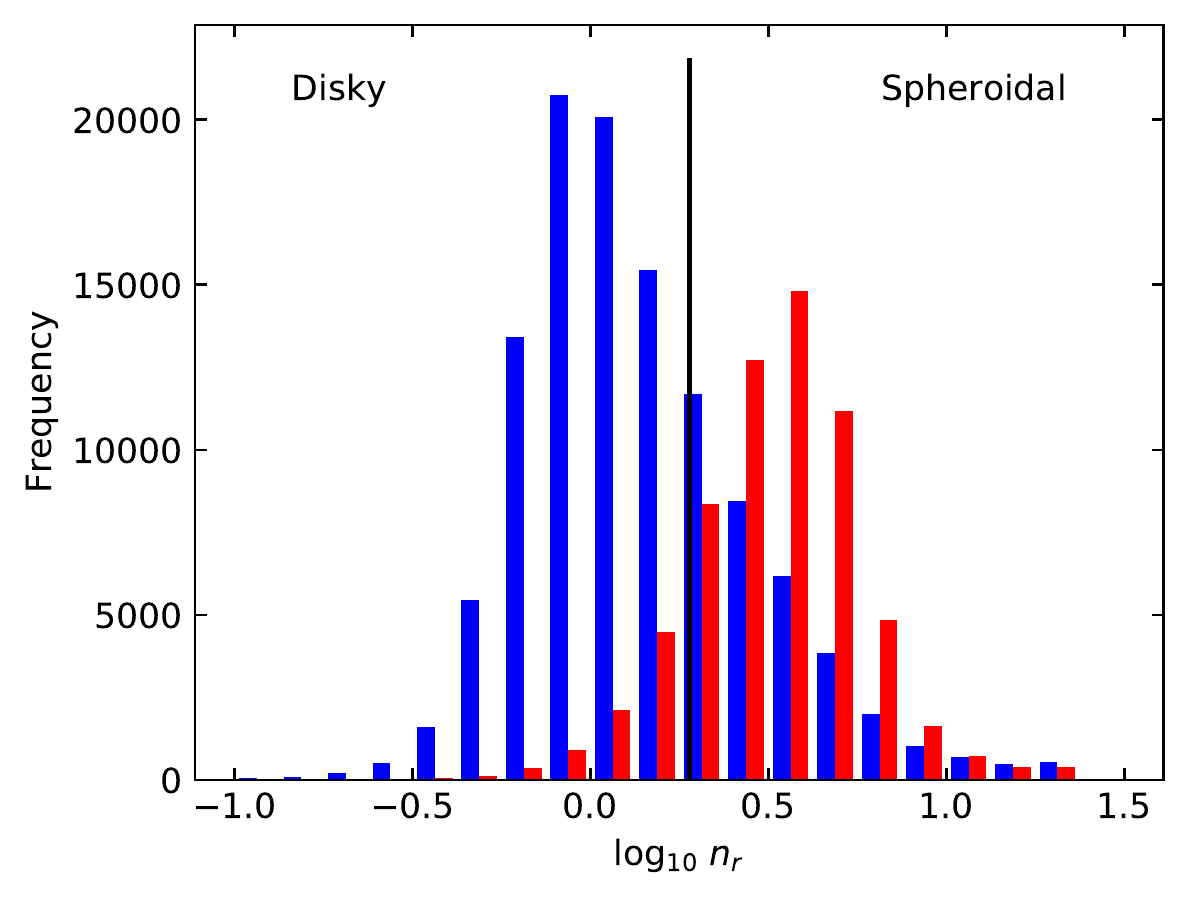} 
\caption{Histogram of GAMA-II log $r$-band \Sersic\ index $n_r$ colour-coded by
  classification into blue and red galaxies.
  The vertical black line shows the separation into disky and spheroidal
  at $n_r = 1.9$.
  While the majority of blue and red galaxies lie to the left and right
  of this line, respectively, there are significant numbers of blue galaxies
  with high index, and vice-versa.}
\label{fig:sersic_hist}
\end{figure}

The LF \citep{Kelvin2014} and SMF \citep{Kelvin2014a,Moffett2016} 
have been presented for galaxies separated into
five bins of morphological type using the GAMA {\tt VisualMorphology} DMU.
However, these visual morphologies are only available for a very local sample
($z < 0.06$).
Many techniques have been developed to make an objective
classification and also to classify thousands of galaxies automatically
(e.g. \citealt{Huertas-Company2015}); however, these methods work well
only with highly-resolved images. At the moment, GAMA does not have
images with sufficient resolution at $z \ga 0.15$. Simple methods, using the
\Sersic\ index \citep{Sersic1963}, give a reliable classification at
least to distinguish between spheroidal and disk-dominated galaxies
(e.g. \citealt{Barden2005}). Therefore, we have made a simple
classification based on the $r$-band \Sersic\ index, $n_r$,
taken from the GAMA DMU {\tt SersicCatSDSSv09} \citep{Kelvin2012}.
Galaxies are considered as spheroidal (or high-$n$)
when $n_r > 1.9$ and disky (or low-$n$) when $n_r < 1.9$. 
Many authors take the cut to be 2.5 (e.g. \citealt{Barden2005});
however, \cite{Kelvin2012} show in their Fig. 15 that the GAMA \Sersic\
index distribution in the $r$-band is bi-modal, with a minimum at $n_r = 1.9$.
We show a histogram of log $r$-band \Sersic\ index colour-coded by
classification into blue and red galaxies in Fig.~\ref{fig:sersic_hist}.
While the majority of blue and red galaxies correspond to disky and spheroidal
respectively, there are significant numbers of blue galaxies
with high index, and vice-versa.

\begin{figure} 
\includegraphics[width=\linewidth]{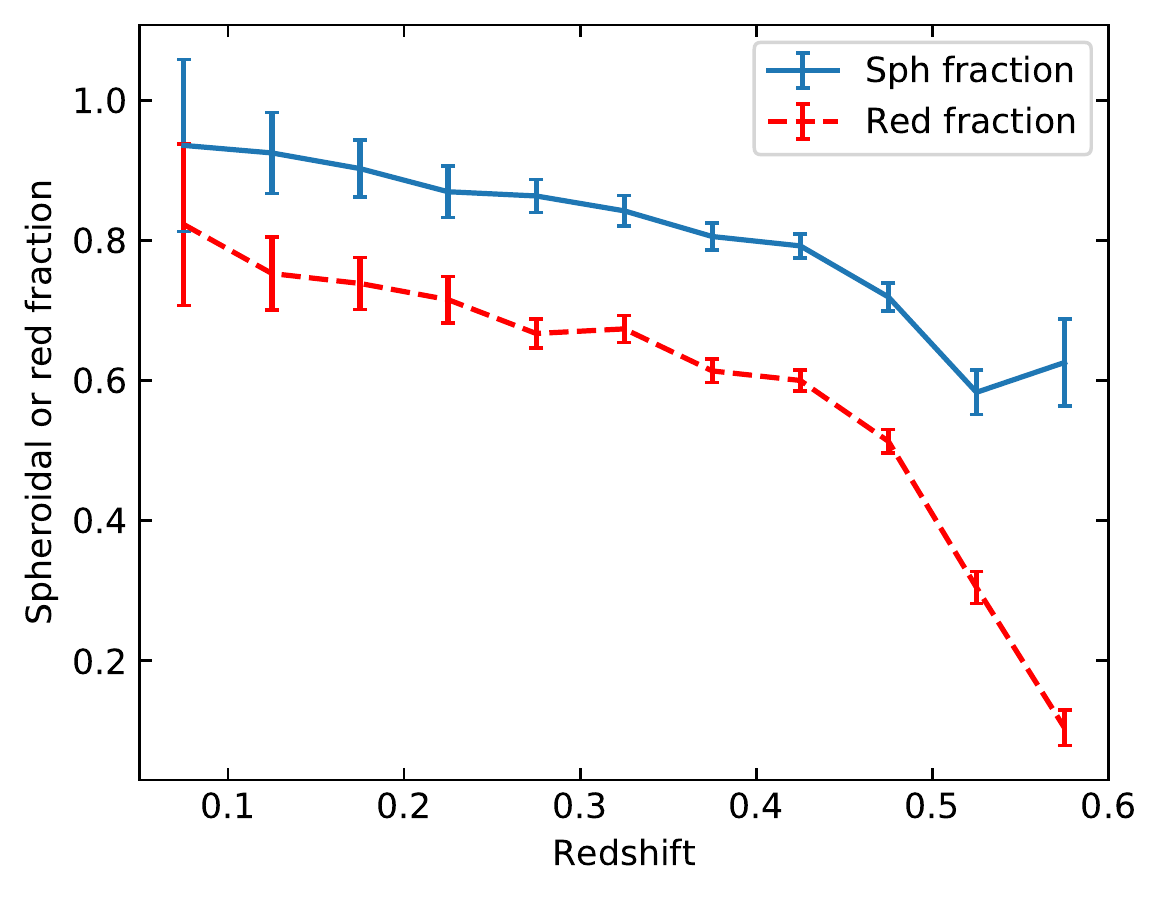} 
\caption{Fraction of luminous ($-22.5 \le M_r < -21.5$) field galaxies
  classified as spheroidal (blue solid line)
  or red (red dashed line) in $\Delta z = 0.05$ bins of redshift.
  For $z < 0.5$, one sees that the spheroidal fraction
  closely tracks the red fraction, thus suggesting any bias in measured
  \Sersic\ index with redshift is minimal.}
\label{fig:sph_red_fraction}
\end{figure}

While the GAMA \Sersic\ modelling takes account of the image
point spread function,
one might still worry that galaxies observed at higher redshift are
less likely to be resolved in SDSS imaging, and thus might have their
\Sersic\ indices biased low (a Gaussian profile corresponds to $n = 0.5$).
To test for this, in Fig.~\ref{fig:sph_red_fraction} we plot the fractions
of luminous ($-22.5 \le M_r < -21.5$)\footnote{Without applying these
  luminosity limits, the red and spheroidal fraction both strongly
  increase with redshift, since high-redshift galaxies tend to be more
  luminous in a flux-limited sample.  We choose to show luminous galaxies
  since this is where we see domination by spheroidal systems in the group LFs.}
field galaxies classified as either spheroidal or as red by our above cuts,
in $\Delta z = 0.05$ bins of redshift.
For $z < 0.5$, corresponding to the redshift limit of our group sample,
one sees that the spheroidal fraction closely tracks the red fraction,
thus suggesting any bias in \Sersic\ index with redshift is minimal.

\subsubsection{Completeness}

\citet{Loveday2012} discuss three sources of incompleteness in GAMA-I data:
incompleteness in the SDSS input catalogue
(primarily a function of surface brightness),
incompleteness in GAMA target selection, and redshift failures.
For the $r$-band LF, target completeness is essentially 100\%
\citep{Loveday2012}.
Therefore, we correct only for input catalogue incompleteness
and redshift failures, following the GAMA-II updates of \citet{Loveday2015}.

GAMA sample selection is complete in $r$-band magnitude,
but not in stellar mass ---
blue galaxies are visible to higher redshifts than red galaxies.
We determine stellar mass completeness as a function of redshift following
a simplified version of the method described in Appendix C of
\citet{Wright2017}.
One would expect the SMF to keep rising to lower masses (at least down to
$\lg M_* \sim 8$ or so), and so we estimate mass completeness by locating
the turn-over point in stellar mass density as a function of redshift.

\begin{figure} 
\includegraphics[width=\linewidth]{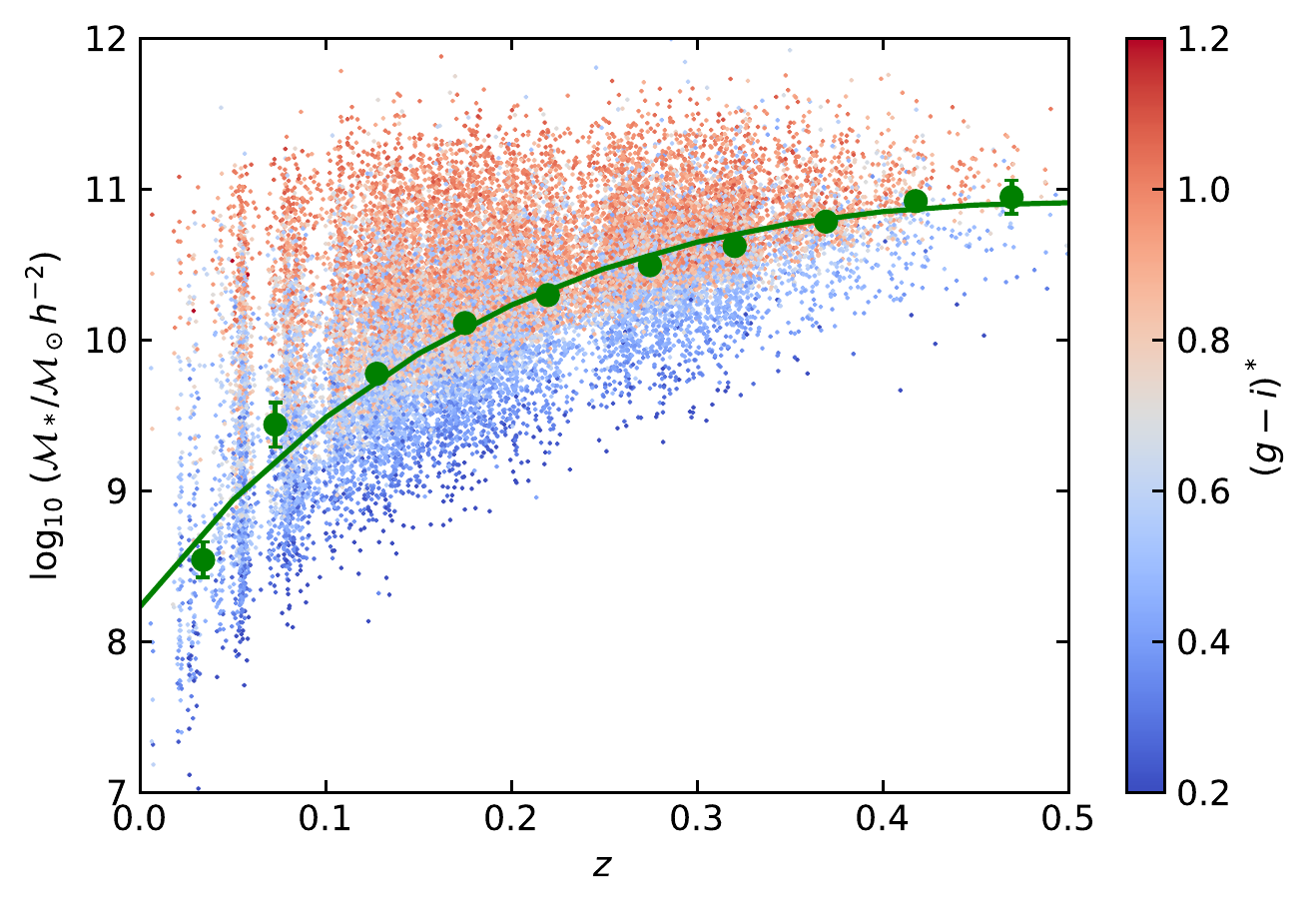}
\caption{Scatter-plot of galaxy stellar mass against redshift
  for grouped GAMA galaxies.
  Galaxies are colour-coded according to intrinsic $(g-i)^*$ colour
  as indicated.
  Large symbols indicate the turnover point in log stellar mass density
  $\lg \mass_*^t$ and its standard deviation in bins of redshift.
  The line shows a second-order polynomial best-fit relation between
  $\lg \mass_*^t$ and scale factor $a = 1/(1+z)$.
  }
\label{fig:gal_mass_z}
\end{figure}

Fig.~\ref{fig:gal_mass_z} shows a scatter-plot of log galaxy stellar mass
against redshift for our sample of grouped GAMA galaxies.
Galaxies are colour-coded according to intrinsic $(g-i)^*$ colour
as indicated.
We consider ten equally-spaced bins in redshift,
ranging from $z=0.0$ to $z=0.5$.
Within each redshift bin, we determine the kernel density estimate (KDE) of
$\lg \mass_*$, using a Gaussian smoothing kernel and default bandwidth
as determined by the routine \verb|scipy.stats.gaussian_kde|.
The turn-over point in stellar mass, $\lg \mass_*^t$, is then chosen as the
maximum of the KDE.
Uncertainty in $\lg \mass_*^t$ is estimated by recalculating the KDE for 100
bootstrap samples of the $\lg \mass_*$ data in each redshift bin.
These turn-over points and uncertainties are indicated by the large symbols
with error bars.
Finally, we fit a second-order polynomial to $\lg \mass_*^t$ as a function
of scale factor\footnote{We found that a quadratic function provides a
  better fit to scale factor than to redshift.} $a = 1/(1+z)$.
We do \emph{not} inverse-variance weight the $\lg \mass_*^t$
estimates in this fit, as the very small uncertainties in $\lg \mass_*^t$ at
intermediate redshifts result in over-fitting to intermediate bins and
poor fit behaviour at low and high redshifts.
The polynomial fit, shown by the curve, is given by
\begin{equation}  \label{eqn:mass_comp}
\lg \mass_*^t = 1.17 + 29.69 a - 22.58 a^2.
\end{equation}
SMF estimates include only galaxies above this mass limit;
equation (\ref{eqn:mass_comp}) is also used to determine the visibility
of a galaxy of given stellar mass in the SMF estimate
(see section~\ref{sec:method}).

\subsection{SMF comparison  simulations} \label{sec:sims}

\input sim_group_mass.tex

We compare our GAMA grouped SMF results with predictions from the \lgal\
semi-analytic model \citep[SAM,][]{Henriques2015} and from two recent
hydrodynamical simulations EAGLE and Illustris TNG.
For all three models/simulations, we utilise data cubes at single snapshot
redshifts corresponding roughly to the mean redshift of the GAMA data,
$\bar{z} \approx 0.2$, rather than attempting to create mock light-cones.
This results in a much higher abundance of low-mass haloes than observed
in GAMA data, and so we set halo mass bin limits to give approximately the
same mean halo mass as for the GAMA data,
see Table~\ref{tab:sim_group_mass_def}.

For the \lgal\ SAM, which is based on the Millennium \citep{Springel2005}
and  Millennium-II \citep{Boylan-Kolchin2009} N-body simulations,
we select the closest redshift snapshot
to the mean GAMA redshift individually for each halo bin.
Halo mass is defined by the mass within an overdensity of 200 times the
critical density.

From the EAGLE suite of simulations \citep{Crain2015,Schaye2015}
we utilise snapshot 26 ($z=0.18$) from the largest-volume simulation,
Ref-L0100N1504.
We use \verb|Group_M_Mean200| from the {\tt FOF} table for halo mass,
and \verb|Mass_Star| from the 30 kpc \verb|Aperture| table, for stellar mass;
see \citet{McAlpine2016} for a complete description of the EAGLE database.

From the suite of IllustrisTNG hydrodynamical simulations
\citep{Marinacci2018,Naiman2018,Nelson2018,Nelson2019,Pillepich2018,Springel2018},
we use the full-resolution simulation with the largest box-size
of 300 Mpc ($205 \hMpc$ for $h = 0.6774$), TNG300-1, at redshift $z=0.2$
(snapshot 84).
Halo masses are given by the FoF halo parameter \verb|Group_M_Mean200|,
and stellar masses are obtained from the subhalo parameter
\verb|SubhaloMassInRadType| for particle type 4 (stars),
the stellar masss within twice the stellar half mass radius.
As recommended by \citet[A1]{Pillepich2018},
we multiply the given stellar masses by a resolution correction factor of 1.4,
appropriate for haloes in the mass range $12 < \lg \mass_h < 15$.

We subdivide the \tng\ galaxies into blue and red using the colour cut
\begin{equation}  \label{eqn:tng_colour_cut}
  (g - i)^* = 0.07 \log_{10} (\mass_*/\mass_\odot h^{-2}) + 0.24,
\end{equation}
where $(g - i)^*$ is the intrinsic stellar colour determined from the
subhalo parameters \verb|SubhaloStellarPhotometrics|,
and $\mass_*$ is the resolution-corrected stellar mass.
This is the same as equation (\ref{eqn:colour_cut}) used to select
blue and red GAMA galaxies, except that we have adjusted the zero-point
offset, so that equation (\ref{eqn:tng_colour_cut}) better follows the
`green valley' in \tng\ galaxy colours.

In order to assess the consistency of these simulations with GAMA data,
we compare field (i.e. group-independent) LFs and SMFs in
Appendix~\ref{sec:field_comp}.
We find that the \tng\ LF under-predicts the numbers of low- and
high-luminosity galaxies.
SMFs are in better agreement, although \tng\ over-predicts
the numbers of very massive ($\lg \mass_* \ga 11$) galaxies.

\section{Methods} \label{sec:method}

In this section we describe our methods for estimating the LF and SMF
from GAMA data and mock catalogues;
these estimates are trivial for the simulations,
since they come in the form of volume-limited boxes.
For GAMA data, uncertainties are determined from nine
jackknife samples, each comprising $4 \times 5$ deg of contiguous area.
These yield larger uncertainties than given by assuming Poisson errors.
For mock catalogues, uncertainties come from the scatter between nine
independent realisations.

\subsection{LF and SMF estimators}  \label{sec:vmax_est}

We first determine the limiting redshift $z_{\rm lim}$ of each galaxy
in the sample.
For the LF calculation, $z_{\rm lim} \equiv z_{\rm lim}^{\rm lum}$
is determined by the GAMA survey magnitude limit of $r = 19.8$ mag,
the galaxy's absolute r-band magnitude,
and its redshift-dependent $K-$ and $e-$corrections.
For the SMF calculation,
$z_{\rm lim} = \min(z_{\rm lim}^{\rm lum}, z_{\rm lim}^{\rm mass})$,
where $z_{\rm lim}^{\rm mass}$ is obtained by substituting
the galaxy's mass for $\mass_*^t$ in equation~(\ref{eqn:mass_comp})
and solving for redshift.

We estimate the LFs and SMFs using a density-corrected \Vmax\ estimator,
allowing for the fact that GAMA groups have a minimum membership threshold
of $N_t$ galaxies, where, for this analysis, we have chosen $N_t = 5$.
The limiting redshift $z_{{\rm lim}, j}$ of group $j$ corresponds
to $z_{\rm lim}^{\rm lum}$ of its $N_t$th brightest member:
beyond this redshift the group would drop below the membership
threshold, and hence be excluded from the sample.
Thus the correct limiting redshift to apply to each galaxy $i$ in group $j$
is $z_{{\rm max}, i} = \min(z_{{\rm lim}, i}, z_{{\rm lim}, j})$.
Here $z_{{\rm lim}, i}$ is the limiting redshift of galaxy $i$
determined as described in the first paragraph of this sub-section,
i.e. neglecting the requirement that its host group be selected.

For a sample bounded by redshift limits $(z_{\rm lo}, z_{\rm hi})$,
we weight galaxy $i$ by $1/\Vdcmax_{,i}$, where
\begin{equation}
  \label{eq:vdcmax}
  \Vdcmax_{,i} = \int_{z_{\rm lo}}^{\min(z_{\rm hi},\; z_{{\rm max},i})}
  \Delta(z) P(z) V(z) dz.
\end{equation}
In this equation, $\Delta(z)$ is the relative overdensity
(taken from fits to the entire GAMA-II sample\footnote{In principle,
  one should use $\Delta(z)$ for each sub-sample considered,
  but since these $\Delta(z)$ estimates would be noisy,
  we make the first-order assumption
  that radial overdensities of different samples vary in the same way.}
by \citealt{Loveday2015}),
$P(z) = P(0) 10^{0.4 P_e z}$ parametrizes number density evolution,
and $V(z)$ is the comoving volume element at redshift $z$.
This estimator has been derived by maximum likelihood
\citep{Cole2011,Loveday2015} and provides a straightforward way of
accounting for both density fluctuations and redshift evolution within
the galaxy sample being analysed.

Higher-mass groups tend to be found at higher redshift (Fig.~\ref{fig:mass_z}),
and so to separate the effects of redshift evolution and environment,
we apply evolution corrections parametrized by $Q_e = 1, P_e = 1$
for luminosity and density evolution respectively.
The corrected absolute magnitude is given by
$M_c = M + Q_e z$ and the density evolution parameter $P_e$
is defined in the preceding paragraph (see also \citealt{Lin1999,Loveday2015}).
To first order, these corrections will take out evolutionary effects
so as to isolate the effects of environment on the LF.

To estimate the LF and SMF for a given sample of galaxies, we simply count
galaxies in bins of absolute $r$-band magnitude or $\lg \mass_*$, respectively,
weighting each galaxy by its $1/\Vdcmax$.

Our LF estimator is tested in Appendix~\ref{sec:simtests},
and compared with estimates of the CLF (number of galaxies per group,
rather than per unit volume).
We find that that unbiased LFs may be estimated without applying redshift cuts,
whereas the CLF estimator will overestimate the number of luminous galaxies
unless a volume-limited group sample is defined,
which would severely reduce the sample size.
For this reason, we show only LF and SMF results,
and not their conditional (per-group) variants, the CLF and CSMF.

\subsection{Functional fits}

Following \citet{Yang2008,Yang2009}, we fit log-normal functions to the
LFs and SMFs of central galaxies, and Schechter functions to those
of satellite galaxies.

Explicitly, the log-normal LFs and SMFs take the form
\begin{equation} \label{eqn:logn}
  \phi_c(M) = \phi^*_c \exp\left[-\frac{(M - M_c)^2}{2 \sigma_c^2}\right],
\end{equation}
where $\phi^*_c$, $M_c$ and $\sigma_c$ correspond to the peak height,
central value and standard deviation of the distribution respectively,
and $M$ refers either to magnitude (LF) or log mass (SMF).

Satellite galaxies may be fit by generalised Schechter functions of the form
\begin{equation} \label{eqn:schec}
  \phi_s(L)\ dL = \phi^*_s \left(\frac{L}{L^*} \right)^\alpha
  \exp\left[-\left(\frac{L}{L^*}\right)^\beta \right]
  d \left(\frac{L}{L^*} \right),
\end{equation}
where $L$ is either luminosity (LF) or stellar mass (SMF),
$\phi^*_s$ is the normalisation, $L^*$ the characteristic luminosity or
stellar mass,
and $\alpha$ the faint-end or low-mass slope,
such that $\alpha = -1$ corresponds to fixed number density
per unit magnitude or per unit log-mass.
The parameter $\beta$, the power to which $L/L^*$ is raised within
the exponential, varies the rate at which the function drops at the
bright/high-mass end.
\citet{Yang2008,Yang2009} use $\beta \equiv 2$ to fit their satellite
LFs and SMFs.
We instead use a standard Schechter function, with $\beta \equiv 1$,
since that gives a slightly better fit (smaller $\chi^2$ values) to our results.
While fits are improved further if we allow $\beta$ to vary as a free parameter,
the strong degeneracy between $L^*$ and $\beta$ makes any trends with
halo mass difficult to interpret.

We fit to LFs over the range of absolute magnitudes $-24 <\, ^{0.0}\!M_r < -16$,
and to SMFs over the mass range $9.0 < \lg \mass_* < 12.5$.
While there are some reliable GAMA SMF measurements for $\lg \mass_* < 9.0$,
the simulations, particularly \tng, are not fully resolved below this mass
limit.

When tabulating functional fits, we quote non-marginalized 1-$\sigma$
errors on the parameters.
For likelihood plots of the shape parameters,
we show 1-$\sigma$ likelihood contours,
but now marginalize over the normalisation parameter $\phi^*$.

\subsection{Redshift evolution}

In order to investigate evolution in the LF and SMF,
we subdivide the sample into three redshift slices given by
$z = [0.002, 0.1]$, [0.1, 0.2] and [0.2, 0.3].
From Fig.~\ref{fig:mass_z}, we see that the group catalogue is
approximately complete to redshift $z=0.3$ for groups of mass
$\lg \mass_h \approx 13.7$ and higher --- see also Appendix~\ref{sec:simtests}.
We therefore use only mass bins \mass3 and \mass4 when subdividing by redshift.
Since we are now explicitly isolating evolutionary effects by subdividing
the galaxies into redshift slices, we `switch off' evolution corrections,
that is we set the evolution parameters to $P_e = Q_e = 0$.

When subdividing by redshift, it is necessary to set completeness limits
on the luminosity and mass range on the LF and SMF, respectively,
as discussed in Section~3.3 of \citet{Loveday2012}.
For the LF, we set a faint absolute magnitude limit given by assuming a
$K$-correction at the lower redshift limit corresponding to the
95-th percentile of the subsample under analysis,
thus assuring that the faintest bin used is at least 95 per cent complete.
For the SMF, the stellar mass limit as a function of redshift is determined
from equation (\ref{eqn:mass_comp}).

\section{Results}  \label{sec:results}

\subsection{Group galaxy LF}


\begin{table}
  \caption{Log-normal fits (equation~\ref{eqn:logn}) to the
    central galaxy LF for different galaxy samples as indicated.
    The final column gives the $\chi^2$ value and degrees of
    freedom $\nu$ of each fit; these fits are mostly good.
  }
  \label{tab:clf_gauss}
  \input clf_cen.tbl
\end{table}

\begin{table}
  \caption{Schechter function fits
    (equation~\ref{eqn:schec} with $\beta \equiv 1$)
    to the satellite galaxy LF for different galaxy samples as indicated.
    The final column gives the $\chi^2$ value and degrees of
    freedom $\nu$ of each fit.
  }
  \label{tab:clf_schec}
  \input clf_sat.tbl
\end{table}

\begin{figure*} 
  \includegraphics[width=0.48\linewidth]{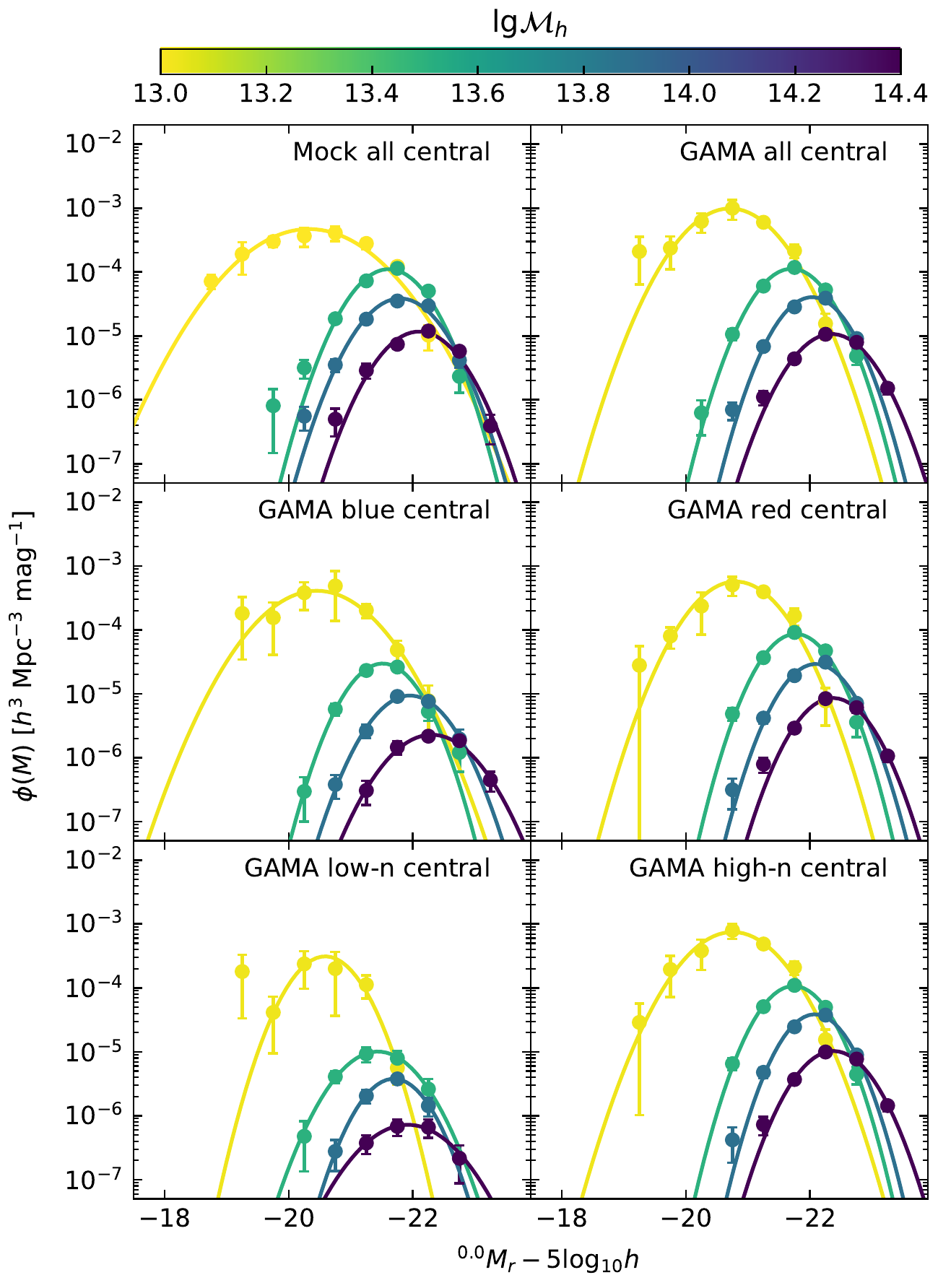}
  \hfill
  \includegraphics[width=0.48\linewidth]{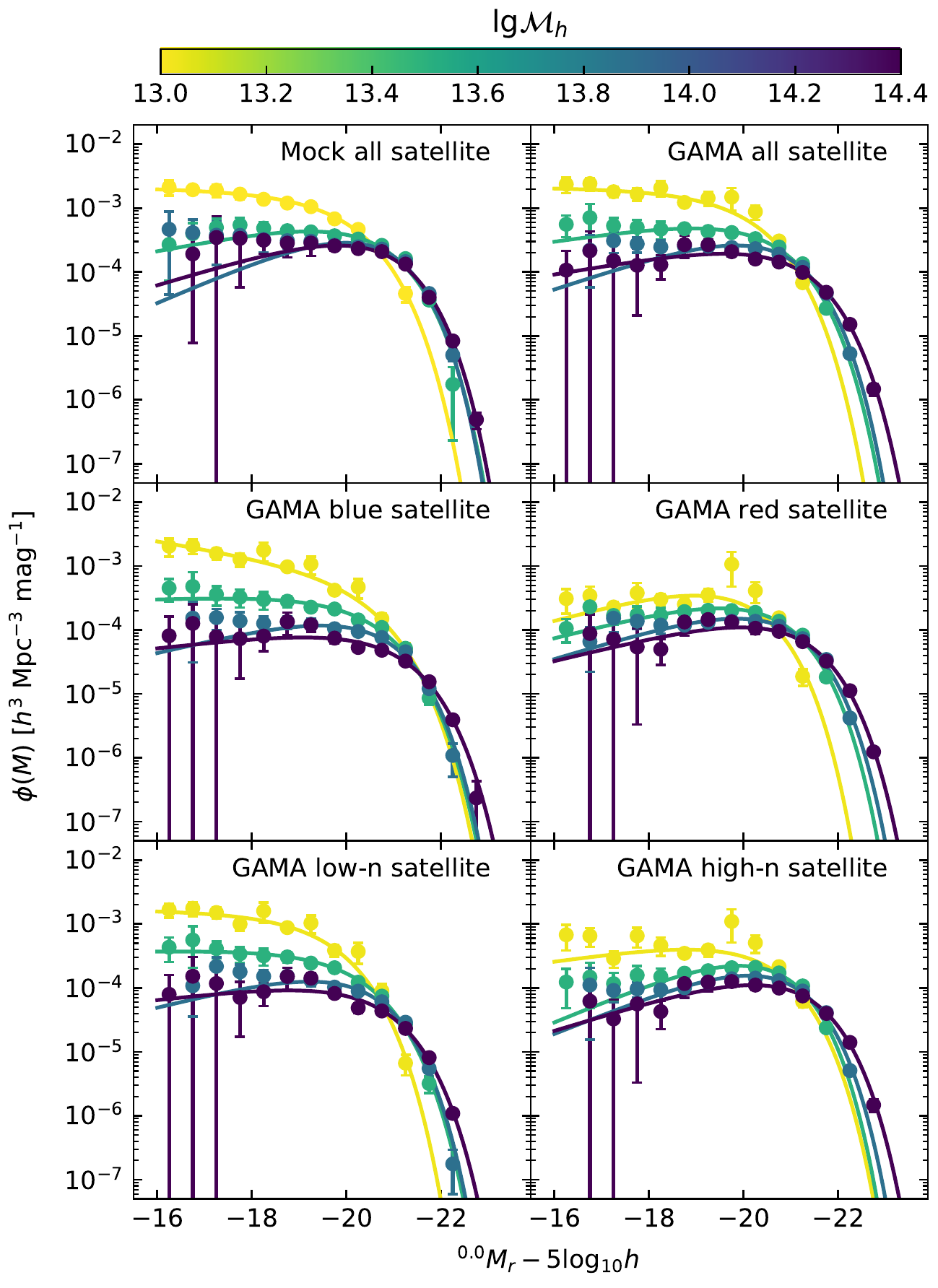}

  \includegraphics[width=0.48\linewidth]{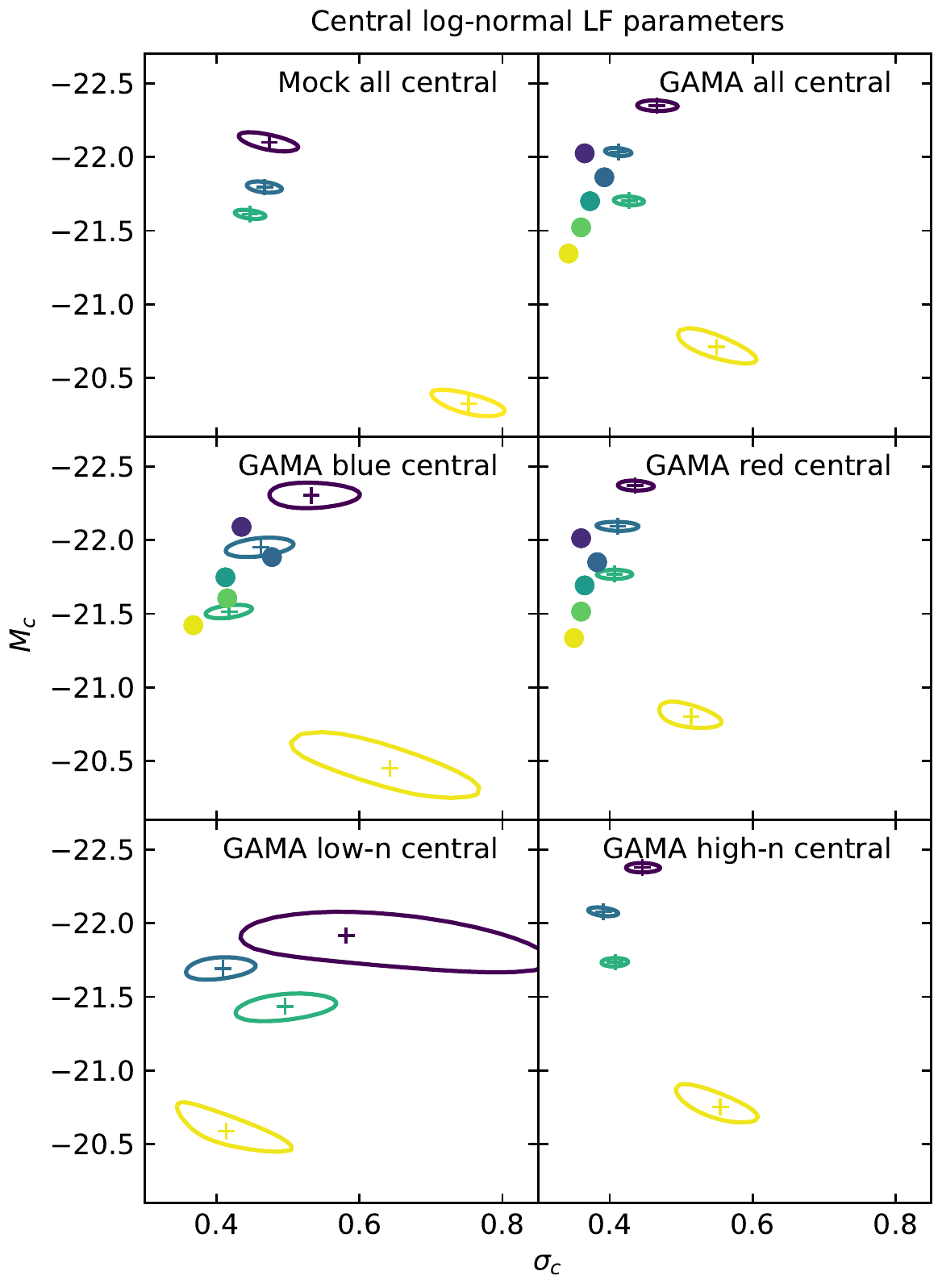} 
  \hfill
  \includegraphics[width=0.48\linewidth]{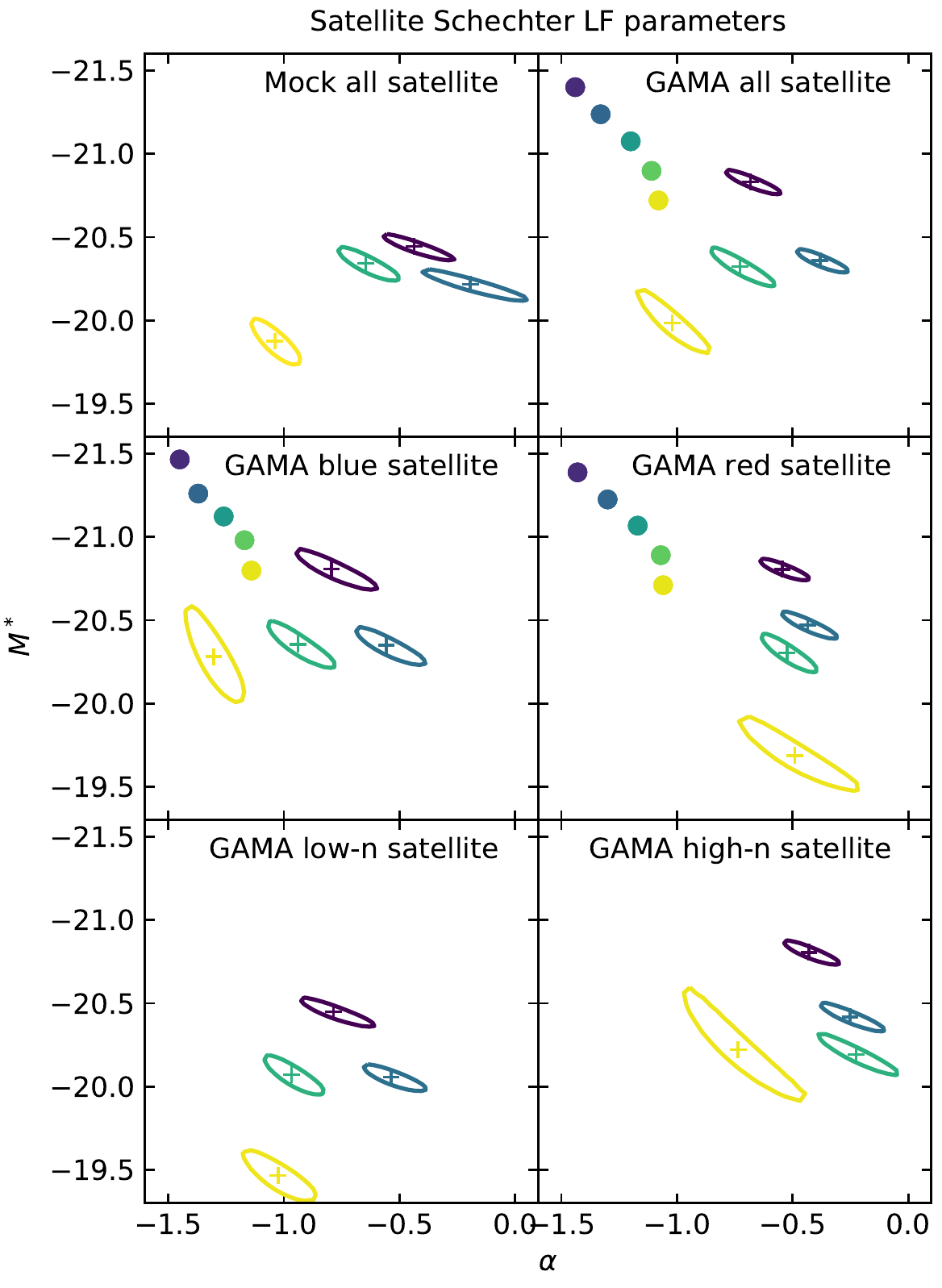} 
 
  \caption{LFs colour-coded by halo mass,
    for central (left-hand panels) and satellite (right-hand panels)
    galaxy samples as labelled.
    Functional fits are log-normal for central galaxies and
    Schechter functions for satellites,
    with 1--$\sigma$ likelihood contours in lower-left and lower-right
    sets of panels, respectively.
    Filled circles in the lower panels show parameter fits from
    \citet{Yang2008}.
  }
\label{fig:clf}
\end{figure*}

Our LF results, colour-coded by halo mass, are plotted in Fig.~\ref{fig:clf}.
Log-normal and Schechter parameter fits for central and satellite
galaxies respectively are
tabulated in Tables~\ref{tab:clf_gauss} and \ref{tab:clf_schec}.



\subsubsection{Central versus satellite}

Fig.~\ref{fig:clf} plots the LFs of central and satellite
galaxies, in the left- and right-hand panel sets, respectively.
Unsurprisingly, central galaxies dominate the bright-end of each LF,
while satellites dominate the faint end.
Due to the trend of increasing group membership with halo mass
(Fig.~\ref{fig:mass_z}), satellite galaxies become an increasingly
dominant contributor to overall group luminosity as halo mass increases.

On the whole, central galaxy LFs are well fit by log-normal functions
(Table~\ref{tab:clf_gauss}), although the mock LFs are slightly skewed to
lower luminosities.
Schechter functions provide generally good fits to the satellite LFs,
although sometimes they under-fit the faint-end in higher-mass groups.

Mock catalogue results
show trends consistent with GAMA, although, as expected from the field
LF comparison in Fig.~\ref{fig:lf_field_comp},
mock central galaxies tend to be offset to slightly
lower luminosity than GAMA centrals, particularly in \mass1 groups.

\subsubsection{Colour and morphology dependence}

The LFs of colour- and \Sersic\ index-selected galaxies
show similar behaviour.
Within halo-mass bins, the central galaxy peak magnitude $M_c$ and satellite
galaxy characteristic magnitude $M^*$
show remarkably little variation with galaxy colour (with the exception
of \mass1 groups, in which blue galaxies are fainter in $M_c$,
but brighter in $M^*$),
whereas spheroidal galaxies tend to be brighter than disky galaxies.
Relative to blue and disky galaxies,
red and spheroidal galaxies are offset to a shallower (more positive)
faint-end slope $\alpha$.

We see that red, and particularly spheroidal,
galaxies dominate the central population, particularly at high halo masses.
The spheroidal/disky ratio of centrals is larger than the red/blue ratio,
particularly in higher-mass haloes.

\subsubsection{LF parameter trends with halo mass}

For central galaxies (lower-left panels of Fig.~\ref{fig:clf}),
we see that peak magnitude $M_c$ brightens systematically with halo mass.
The width of the magnitude distribution $\sigma_c$ is largely independent
of halo mass, although is broader in the lowest-mass haloes.

Within each satellite galaxy class we observe
(lower-right panels of Fig.~\ref{fig:clf})
a systematic and significant brightening of the
characteristic magnitude $M^*$ with increasing halo mass.
Any trends of faint-end slope $\alpha$ are less clear,
although for most samples, galaxies in \mass1 haloes show the
steepest faint-end slope.
Mock galaxies show consistent trends with the `GAMA all' sample.

For comparison, we also show, in the lower panels of Fig.~\ref{fig:clf},
log-normal and modified ($\beta \equiv 2$) Schechter function fits
to the central and satellite populations respectively from \citet{Yang2008}
as filled circles (they do not split galaxies by morphology).
We use parameter values from Table~1 of \citet{Yang2008} for their
five halo mass bins within the range $13 \le \lg \mass_h < 14.4$.
Note that as well as the difference in satellite fitting function,
\citet{Yang2008} $K$-correct to $z=0.1$ rather than $z=0.0$,
and use a different colour cut,
but one would nevertheless hope that trends with halo mass would be preserved.

For central galaxies, we observe consistent, but more pronounced,
trends of $M_c$ with halo mass, cf. \citet{Yang2008}.
This difference could be explained by
underestimated halo masses in \citet{Yang2007} single-galaxy groups
(see fig.~6 of \citealt{Davies2019}),
and so the \citet{Yang2008} low-mass bin likely mixes haloes of
both low and high mass.

For satellite galaxies, the \citet{Yang2008} characteristic magnitudes $M^*$
and faint-end slopes $\alpha$
respectively are offset to significantly brighter and steeper values than ours,
an effect attributable to the different choice of fitting function.
Their observed trend of brightening $M^*$ with halo mass is consistent with ours.
Contrary to our results, they see a clear steepening of faint-end slope $\alpha$
with increasing halo mass.
One should note, however, that there is a hint in 
\citet[][fig.~2]{Yang2008} that the faint-end slope may be a little
too shallow cf. their non-parametric estimates in lower-mass haloes.

\subsection{Group galaxy  SMF}

\begin{figure*} 
  \includegraphics[width=0.48\linewidth]{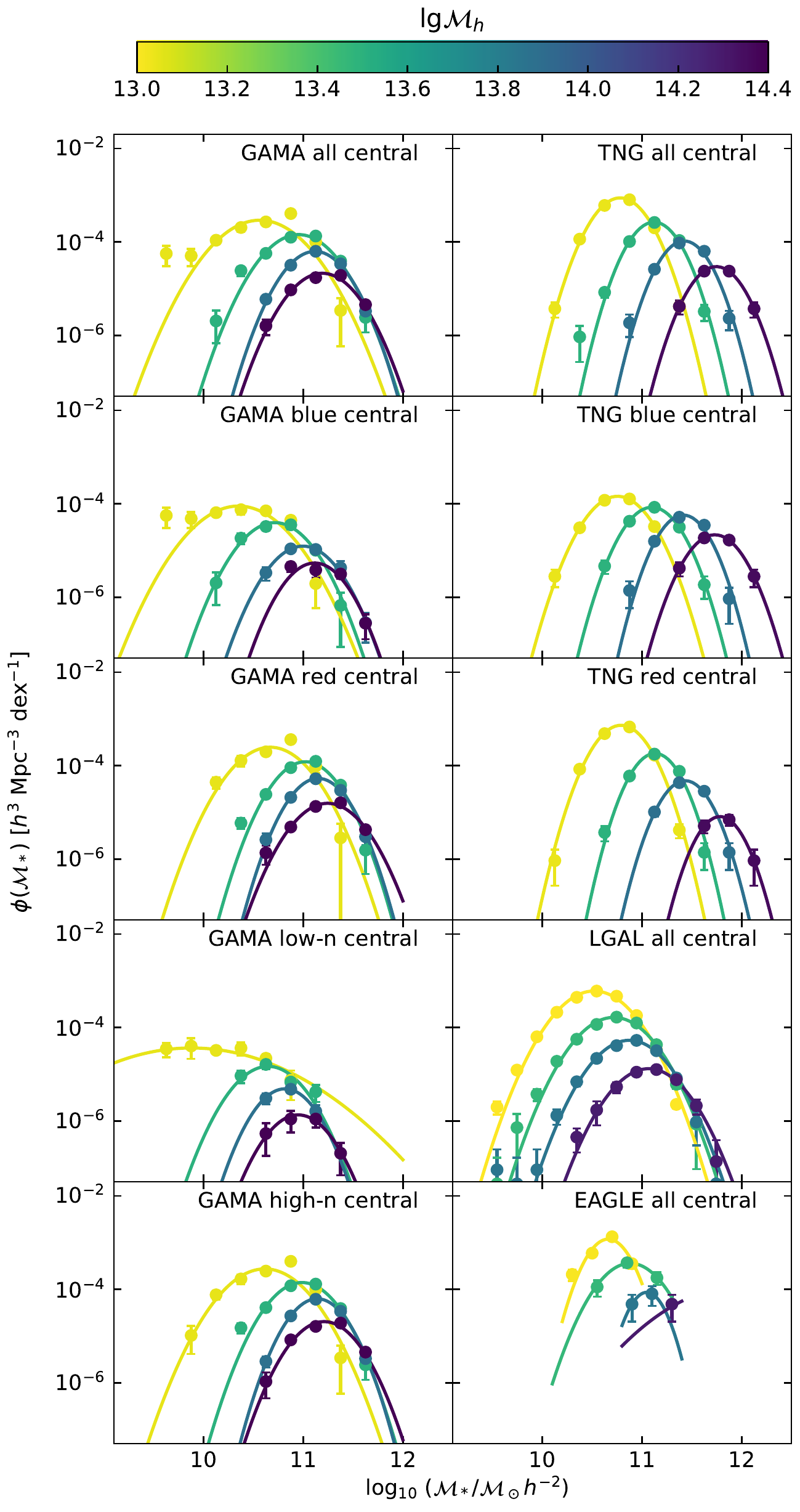} 
  \hfill
  \includegraphics[width=0.48\linewidth]{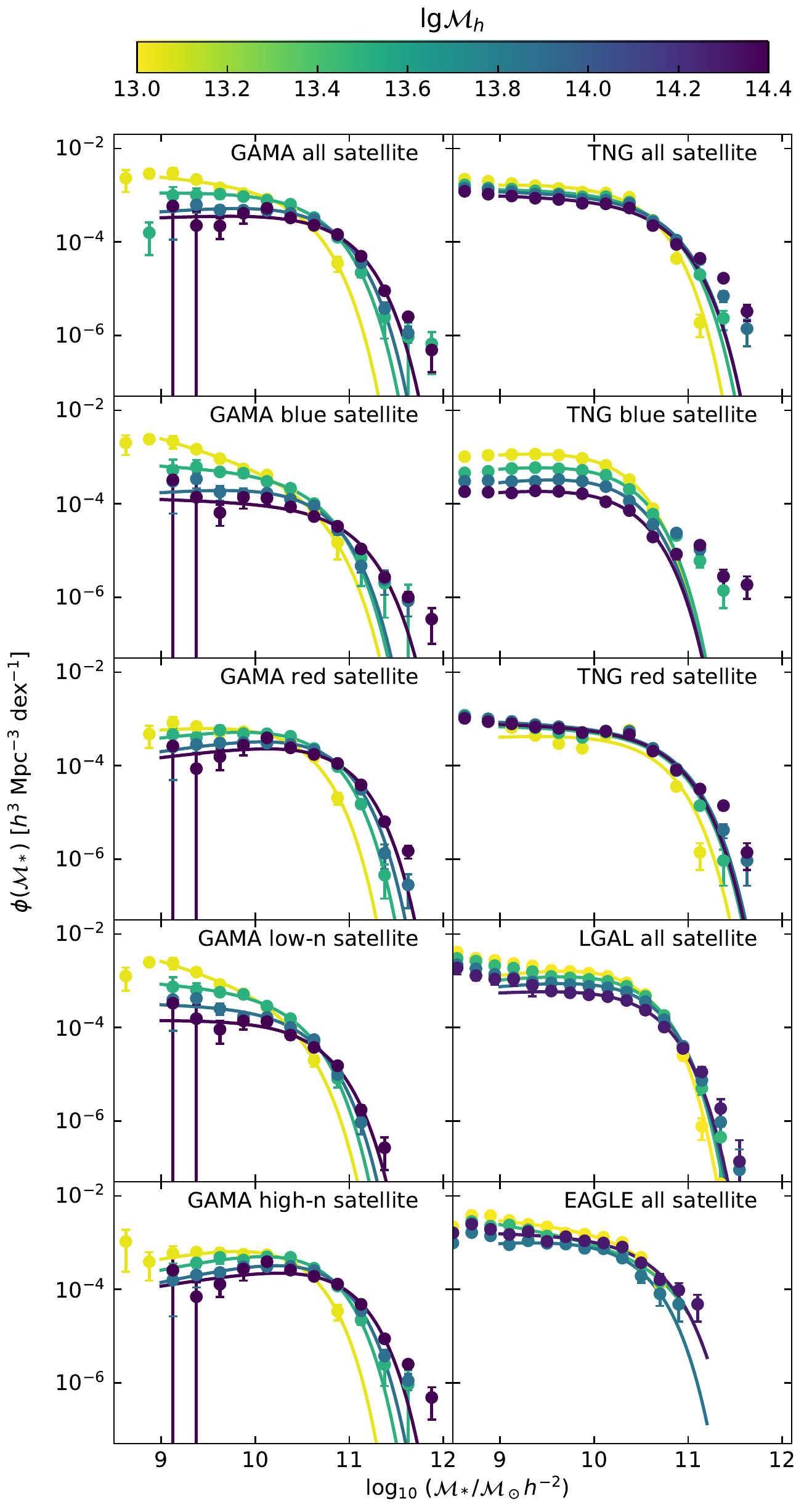} 
  \caption{SMFs colour-coded by halo mass,
    for galaxy samples as labelled in each panel.
    Central and satellite galaxies are shown in left- and right-hand panel sets,
    and fitted with log-normal and Schechter functions, respectively.
  }
\label{fig:csmf}
\end{figure*}

\begin{figure*} 
  \includegraphics[width=0.48\linewidth]{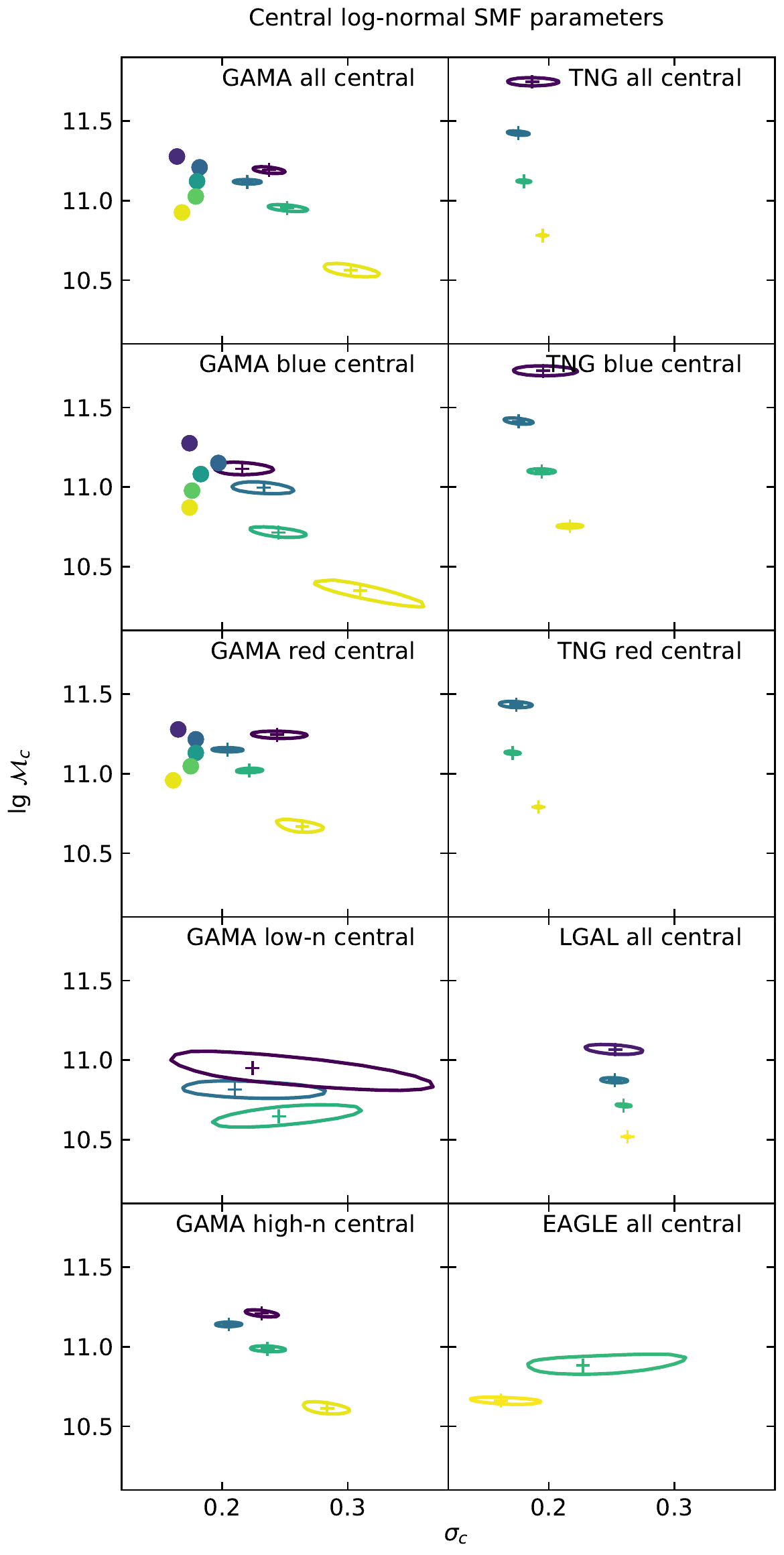} 
  \hfill
  \includegraphics[width=0.48\linewidth]{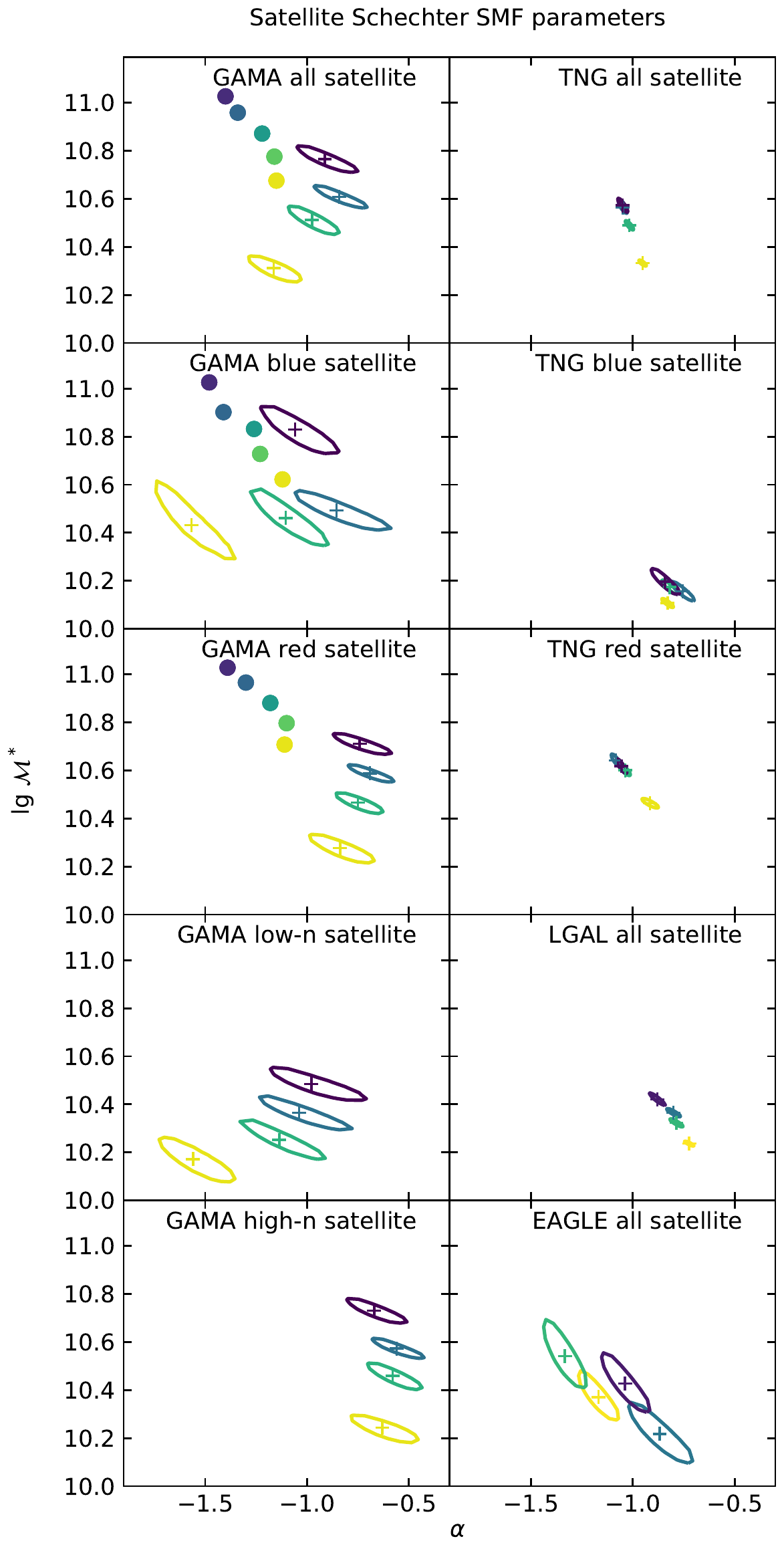} 
 
  \caption{1-$\sigma$ likelihood contours for log-normal fits to
    central galaxies (left) and Schechter function
    parameter fits to satellite SMFs (right), colour-coded by halo mass
    using the same scheme as Fig.~\ref{fig:csmf}.
    Filled circles show parameter fits from \citet{Yang2009}.
  }
\label{fig:csmf_fit}
\end{figure*}

Our SMF results, along with those from the \lgal\ SAM,
and the EAGLE and \tng\ simulations, are plotted in Fig.~\ref{fig:csmf}.
Note that the relative normalisation of GAMA data and simulations is
somewhat arbitrary, depending as it does on the halo mass limits.
Log-normal and Schechter parameter fits for central and satellite
galaxies respectively are shown in Fig.~\ref{fig:csmf_fit} and
tabulated in Tables~\ref{tab:csmf_gauss} and \ref{tab:csmf_schec}.
We first discuss the observed SMF results for GAMA galaxies subdivided by
central/satellite, colour and morphology,
comparing with SDSS results from \citet{Yang2009}.
We then compare observed results with those from the \lgal\ SAM and simulations.

\subsubsection{Observed central versus satellite}

Fig.~\ref{fig:csmf} plots the SMFs of central
and satellite galaxies in the left- and right-hand panel sets, respectively.
Unsurprisingly, central galaxies dominate at high stellar mass,
while satellites dominate at low mass.
As with the LFs, satellites become more dominant in high-mass haloes due to
the mass--richness correlation for groups.

On the whole, central galaxy SMFs are reasonably fit by log-normal functions
(Table~\ref{tab:csmf_gauss}), although there are some statistically poor
fits in the lower halo mass bins, due to a slight excess over the
log-normal fit at lower masses.
Schechter functions provide variable-quality fits to satellite SMFs
(Table~\ref{tab:csmf_schec}); in particular we observe a high-mass excess above
the Schechter fit in higher-mass haloes.
One can obtain a better fit by allowing the parameter $\beta$ in equation
(\ref{eqn:schec}) to vary freely.
However, the values of $\beta$ and $M^*$ are strongly correlated,
and so parameter trends with halo mass are much harder to interpret,
and also to compare with previous results.
We thus choose to show only standard Schechter function fits ($\beta \equiv 1$).

\subsubsection{Observed colour and morphology dependence}

At all halo masses, we see that red, and particularly spheroidal,
galaxies dominate the central population.
As with the LFs, the spheroidal/disky ratio of centrals is larger than the
red/blue ratio.
Our morphology-dependent results for low-mass haloes are qualitatively
consistent with the field SMF results of \citet{Moffett2016},
in which spheroidal and disky galaxies dominate at high and low stellar masses,
respectively.

The SMFs of colour- and \Sersic\ index-selected galaxies
show some subtle differences.
For centrals, peak log-mass $M_c$ tends to be higher for red and spheroidal
than for blue and disky galaxies; $M_c$ is particularly low for disky galaxies
in \mass1 haloes.
There are no significant differences in width parameter $\sigma_c$ apart
from a broadening in \mass1 haloes, again particularly for disky galaxies.
For satellites, spheroidal galaxies exhibit higher characteristic stellar mass
$\mass^*$ and shallower low-mass slope $ \alpha$ than disky galaxies,
whereas red and blue galaxies have more consistent SMF shapes,
with the exception of steep low-mass slopes for blue galaxies in \mass1 haloes.

\begin{table}
  \caption{Log-normal fits (equation~\ref{eqn:logn}) to the
    central galaxy SMF for different galaxy samples as indicated.
    The column headed $\chi^2/\nu$ gives the $\chi^2$ value and degrees of
    freedom for the functional fit.
  }
  \label{tab:csmf_gauss}
  \input csmf_comp_cen.tbl
\end{table}

\begin{table}
  \caption{Schechter function fits
    (equation~\ref{eqn:schec} with $\beta \equiv 1$)
    to the satellite galaxy SMF for different galaxy samples as indicated.
    The column headed $\chi^2/\nu$ gives the $\chi^2$ value and degrees of
    freedom for the functional fit.
  }
  \label{tab:csmf_schec}
  \input csmf_comp_sat.tbl
\end{table}

\subsubsection{Observed SMF parameter trends with halo mass}

For central galaxies (left panels of Fig.~\ref{fig:csmf_fit}),
we see that peak log-mass $M_c$ increases systematically with halo mass,
and is $\sim 0.2$ dex higher for red and spheroidal galaxies than their
blue and disky counterparts.
The width of the mass distribution $\sigma_c$ tends to increase for
lower halo masses, particularly for disky galaxies, whose \mass1
likelihood contour lies well off the bottom-right limits of the plot.

Within each satellite galaxy class we observe
(right panels of Fig.~\ref{fig:csmf_fit}) a systematic increase in 
characteristic mass $M^*$ with increasing halo mass.
There is little significant trend of low-mass slope $\alpha$ with halo mass,
except that it is much steeper for blue and disky galaxies
in \mass1 haloes.

For comparison, we also show log-normal and modified ($\beta \equiv 2$)
Schechter function fits
to the central and satellite populations, respectively, from \citet{Yang2009}
as filled circles (they do not split galaxies by morphology).
We use parameter values from Table~4 of \citet{Yang2009} for their five
halo mass bins within the range $13 \le \lg \mass_h < 14.4$.
Note that the \citet{Yang2009} colour-cut is different to ours,
but trends with halo mass should not be strongly affected.
As with the LFs, the \citet{Yang2009} satellite SMF parameters are offset to
brighter and steeper values than ours, due to the different choice of power
within the Schechter function exponential.
We observe consistent trends in peak and characteristic stellar mass
with halo mass for central and satellite galaxies, respectively,
although our halo mass dependence is slightly stronger.
Again, this is likely to be due to the \citet{Yang2009} low-mass bins containing
a range of halo masses.
\citet{Yang2009} find narrower log-normal fits to centrals, possibly reflecting
their narrower bins in halo mass.
Unlike \citet{Yang2009}, we do not observe a systematic steepening of satellite
low-mass slope $\alpha$ with halo mass.
Again, we note \citep[][fig.~4]{Yang2009} that their low-mass slopes in
low-mass haloes may be a little too shallow.

\subsubsection{Comparison of GAMA and simulated SMFs}

For central galaxies, \lgal\ and EAGLE show log-normal parameters and trends
consistent with observations,
with the caveat that the small volume of the EAGLE simulation
(27 $\times$ smaller than TNG-300),
means that there are very few massive galaxies,
hence the central fits are poorly constrained.
\tng\ groups, however, host much more massive central galaxies than the
observations, and show an enhanced dependence of $M_c$ on halo mass.
Central galaxy stellar masses in \tng\ thus seem to be both too high and also
over-dependent on halo mass.

For satellite galaxies,
the functional fits to the large-volume \lgal\ and \tng\ SMFs are often
statistically very poor, with some reduced $\chi^2$ values in excess of 10.
This is partly due to the large numbers of galaxies in these simulations
giving rise to very high signal-to-noise measurements,
but is also due to the fact that Schechter fits 
drop too steeply at the high mass end in massive haloes.
The discrepancy is even worse with the $\beta \equiv 2$ modified Schechter
functions used by \citet{Yang2009}.
One can get a slightly better fit to the \tng\ results by allowing $\beta$
to vary freely, but the $\chi^2$ values are still poor in many cases.
In particular, no value of $\beta$ can match the very shallow
high-mass shape of the \tng\ blue satellite SMF.
The \lgal\ SMFs clearly favour a double Schechter function, with a steeper
slope below mass $\lg \mass_* \approx 9.5$.

The Schechter fits to satellite galaxies show much smaller shape variation
with halo mass than the observations.
SDSS and GAMA observe an increase in characteristic mass of
$\Delta \lg \mass_* \approx 0.4$ and 0.5 dex respectively from \mass1--\mass4
haloes, whereas the simulations yield $\Delta \lg \mass_* < 0.3$ dex.
The \tng\ and \lgal\ SMFs show only a very small steepening of
low-mass slope $\alpha$ with halo mass;
EAGLE shows larger variation in both $\lg \mass^*$ and $\alpha$,
but no systematic trends with halo mass.
We caution that the likelihood contours for simulated satellite galaxies in
Fig.~\ref{fig:csmf_fit} arise from functional fits that are in some cases
statistically very poor.
It can be seen visually in Fig.~\ref{fig:csmf}
that the SMF shapes of low to moderate mass satellite galaxies in the
\lgal\ SAM and the simulations appear to be almost independent of their
host halo mass.
Only for the most massive satellites ($\lg \mass_* \ga 10.5$),
does their abundance increases significantly with halo mass.

\subsection{LF and SMF Evolution}

\begin{figure*} 
\includegraphics[width=\linewidth]{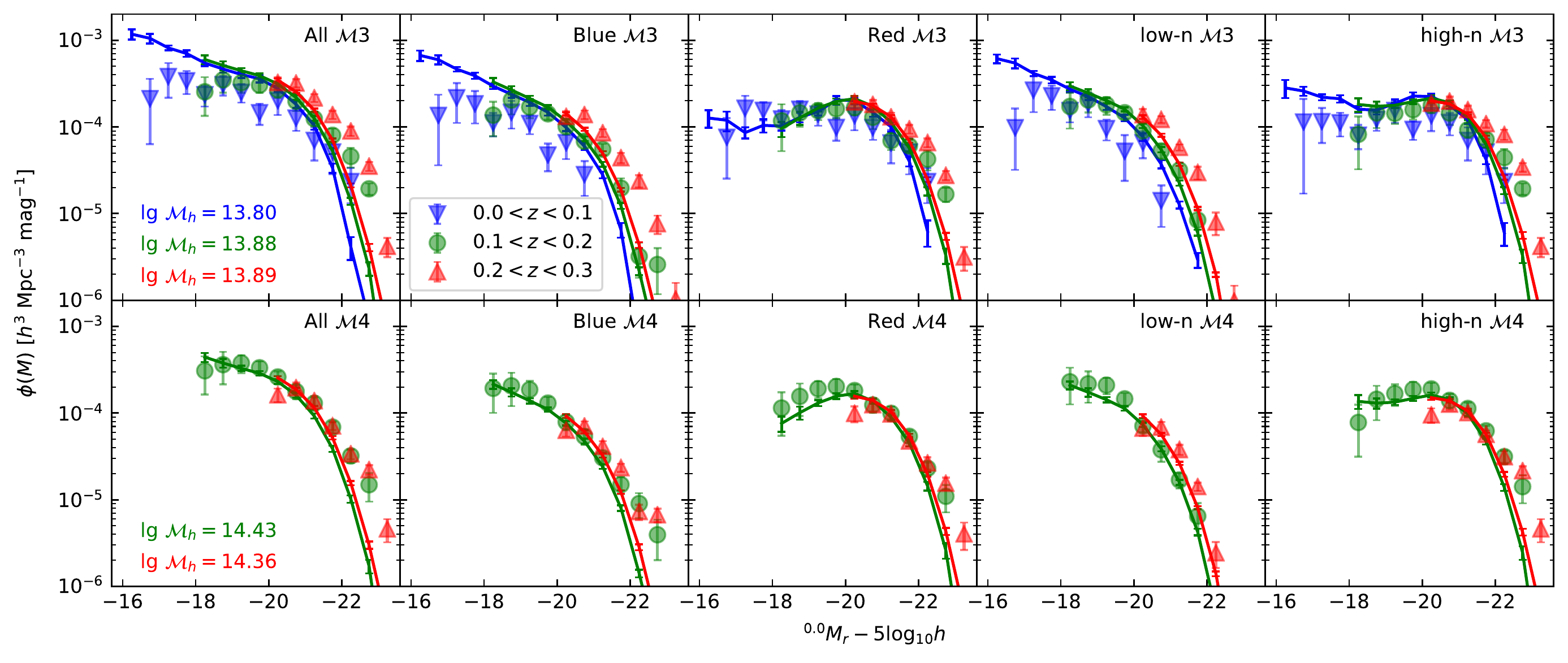} 
\includegraphics[width=\linewidth]{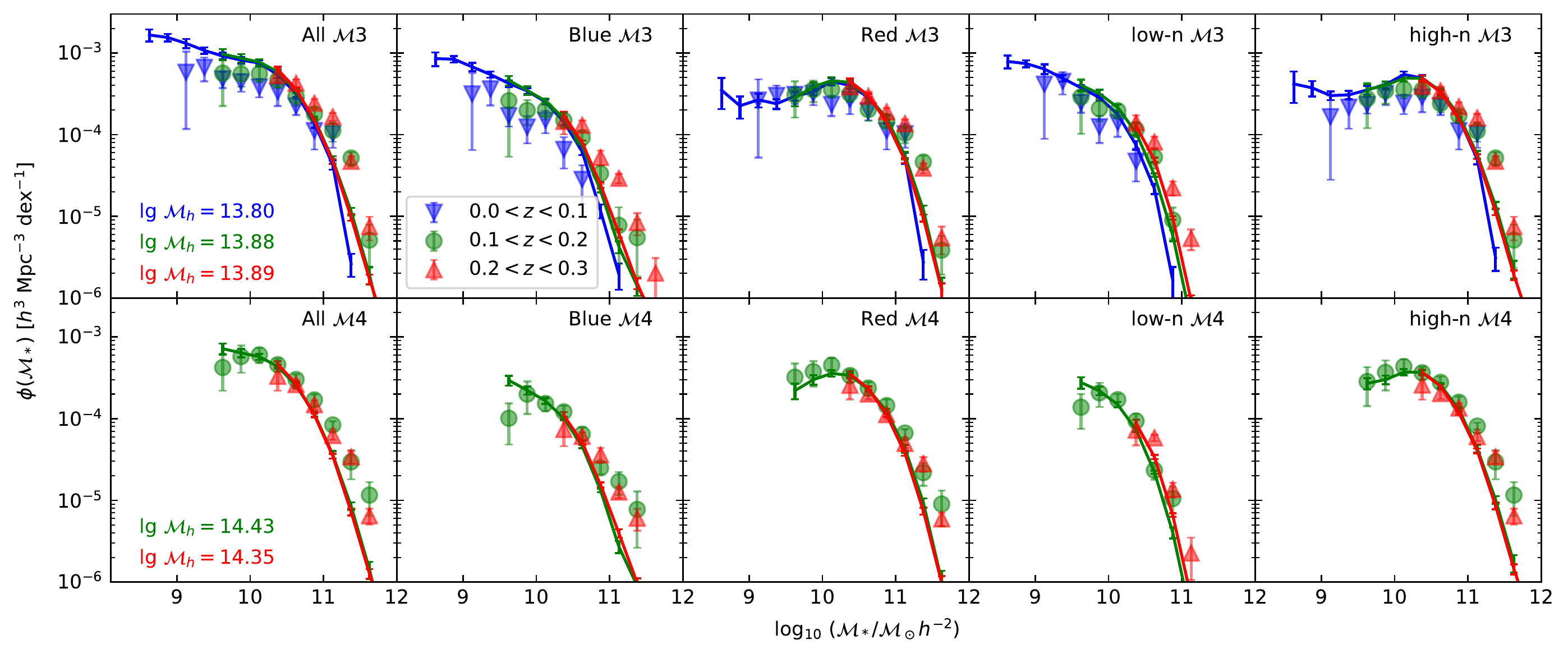} 
\caption{Grouped galaxy LFs and SMFs are shown as symbols in the upper and lower
  set of panels, respectively.
  Within each set of panels, galaxies are subdivided by halo mass (rows)
  and galaxy type (columns), with colour coding and symbol shape
  indicating redshift range.
  The numbers in the left-most panels indicate the mean halo mass within
  each redshift bin, showing that, within halo mass bins, there is
  relatively little dependence of halo mass on redshift.
  For comparison purposes, lines show the LFs or SMFs of field galaxies,
  whether grouped or not, renormalised to the number of grouped galaxies
  in each panel.
  No evolution corrections are applied to the LFs.
}
\label{fig:clf_ev}
\end{figure*}

LFs and SMFs determined in halo mass bins and redshift
slices\footnote{
  We note that the EAGLE halo mass function exhibits negligible evolution
  over the redshift range shown, and so do not expect GAMA halo masses
  to evolve significantly.
}
are plotted in Fig.~\ref{fig:clf_ev}
(upper and lower set of panels, respectively), and show a consistent picture.
Symbols show LFs/SMFs in group mass bins (rows)
and for different galaxy types (columns), as labelled.

For comparison purposes, lines show LFs/SMFs of field galaxies
(i.e. whether grouped or not) of corresponding type,
renormalized to the number of grouped galaxies in each panel
(i.e. of given type and environment, but summing over redshift bins).
For all types of GAMA galaxies, the field LFs
show evidence for minor fading in $M^*$ since redshift $z \approx 0.3$.
The field SMFs show little sign
of evolution, apart from a paucity of the most massive galaxies at low redshift.

In the first column of both panel sets (all galaxies),
we see that the relative number densities of 
luminous ($M_r \la -21$ mag) and massive ($\lg M_* \ga 10.5$) grouped galaxies
are enhanced over the field in both group environments at redshifts $z \ga 0.1$.
In the only redshift bin ($z < 0.1$) in which faint ($M_r \ga -18$ mag)
galaxies are visible, they are relatively less abundant than in the field.
Dwarf galaxies when selected by mass ($\lg M_* \la 9.5$) are more consistent
with the field, but still slightly suppressed.
Overall, the number density of grouped galaxies at low redshift ($z < 0.1$)
is below that in the field, given our renormalization across all redshifts.

Blue and disky galaxies (second and fourth columns) show similar behaviour
to the general population.
Blue galaxies show a slightly larger excess than disky
galaxies over their respective field populations at the bright/high-mass end
at redshifts $z > 0.1$.

Only red galaxies (third column) show a number density consistent with the
renormalized field at low luminosities/masses and low redshift.
This is to be expected if star formation is quenched in the
infalling galaxies, leading to an increased abundance of red galaxies.
Low-luminosity and low-mass red galaxies in \mass4 groups at $0.1 < z < 0.2$
show a slight excess relative to the field.

Spheroidal galaxies (final column) show similar behaviour to the
overall population, with no low-luminosity/mass, low-redshift, excess,
when compared with red galaxies.
This suggests that the quenching process has not yet had time to fully
transform the morphological appearance of infalling galaxies.

Note that, within each halo mass bin, the mean halo mass shows only a weak
dependence on redshift, as indicated by the numbers in the left-most panels
of Fig.~\ref{fig:clf_ev}.
We thus believe that the changes in group LF or SMF with redshift
are primarily due to evolution rather than varying host halo mass.
As a caveat, however,
we note that due to the flux-limited nature of the GAMA sample,
groups of fixed mass have more detected members when observed at lower redshift
(Fig.~\ref{fig:mass_z}),
which may have some effect on the apparent evolution measured.
However, it seems unlikely that this would result in the differences in
behaviour seen for blue and red galaxies.

\section{Conclusions and discussion}  \label{sec:concs}

In this work we have presented the $r$-band LF and the SMF for galaxies
in the GAMA group catalogue (\gtc v9) separated into central and satellite,
and divided by colour and morphology.
The group catalogue was divided into four mass bins \mass1--4 covering
$\lg \mass_h = 12$--15.2 to explore the dependency
of the LF and SMF on group mass, and later into three redshift bins below
$z = 0.3$ to investigate LF/SMF evolution.

On subdividing galaxies into central and satellite populations,
we see that centrals are always luminous and massive,
with fainter ($M_r \ga -19$ mag) galaxies being exclusively satellites,
particularly in higher-mass haloes.
We note that the only (indirect) use of mass in defining a GAMA central galaxy
is the choice of the brighter of the two galaxies that remain after
iteratively rejecting galaxies furthest from the group centre-of-light
(see R11 sec 4.2.1).
This contrasts with, for example, \citet{Yang2008,Yang2009,Knobel2015},
who define the central galaxy as the most luminous or massive in the group.

Due to the correlation between group mass and richness,
satellite galaxies become more and more numerically dominant
in higher-mass groups.
Note that the definition of GAMA groups is very different to that of
\cite{Yang2007}, whose group definition includes those comprising
single galaxies.

When subdividing the galaxy population by colour and \Sersic\ index,
we find that red and, in particular, spheroidal galaxies dominate at
high luminosity and mass;
blue and disky galaxies dominate, or at least contribute roughly equally,
at low luminosity and mass.
The fraction of galaxies classified as red or spheroidal increases with
increasing halo mass, consistent with the findings of \citet{Davies2019}.

We next summarize our results separately for central and satellite galaxies,
and for low-redshift evolution, and then put these findings in context.

\subsection{Central galaxy LF and SMF}

The observed central galaxy LF and SMF are well-fit by log-normal functions,
with peak luminosity and mass increasing systematically with host halo mass.
These trends are consistent with the SDSS measurements by
\citet{Yang2008,Yang2009}, except that we observe a much stronger dependence:
comparing \mass4 and \mass1 haloes, we observe $\Delta M_c \approx 1.6$ mag
or 0.6 dex for the LF and SMF respectively, compared with $\approx 0.7$ mag
and $\approx 0.4$ dex from SDSS.
These differences can be understood if the SDSS halo masses are underestimated
for single-member groups, as indicated by the comparison in
fig.~6 of \citet{Davies2019}.
This would imply that the lower-halo mass bins in \citet{Yang2008,Yang2009}
actually contain haloes with a wide range of masses.

The luminosity and mass distributions tend to be broader in the \mass1 bin,
which covers the widest range of halo masses.
The narrower log-normal fits from \citet{Yang2008,Yang2009} likely
reflect their use of narrower bins in halo mass.

GAMA mock central LFs show consistent trends with the observations,
albeit offset to slightly lower luminosities.
The \lgal\ SAM shows central SMF parameters consistent with GAMA,
whereas the \tng\ simulation yields peak masses both higher than ours,
and with an even more enhanced dependence on halo mass
($\Delta M_c \approx 1.0$ dex).
This suggests that the \tng\ central galaxy stellar masses are overly-dependent
on halo mass.
The small volume of the EAGLE simulation provides only weak constraints
on the halo-mass dependence of the central SMF.

\subsection{Satellite galaxy LF and SMF}

The observed satellite galaxy LF and SMF are reasonably well-fit by standard
Schechter functions, with characteristic luminosity and mass increasing
systematically with host halo mass, in agreement with all previous studies.
Faint-end and low-mass slopes of the satellite LF and SMF, respectively,
show little systematic correlation with halo mass, except that galaxies
in the lowest mass haloes tend to have the most steeply-rising slopes.
This is in disagreement with some previous group LF/SMF results such as
\citet{Yang2008,Yang2009,Robotham2010,Zandivarez2011},
but in agreement with \citet{Phillipps1998},
who find that dwarfs are more common in lower density environments.
\citet[][fig.~15]{Reddick2013} see no clear dependence of CSMF slope
on halo mass for their SDSS group catalogue.
Such discrepancies are likely to arise due to the inability of
Schechter-like functions to accurately match the observed LF shape
over a wide range of luminosities: the faint-end slope
is often more strongly constrained by high signal-to-noise measurements
around $L^*$ than by the faintest galaxies in the sample.
The same argument applies when fitting the SMF.

We note that the dependence of the faint- or low-mass slope
on local density as estimated by galaxy counts
in cylinders or spheres is similarly ambiguous, with some authors 
\citep[e.g.][]{Xia2006,Peng2010} finding a steepening faint-end/low-mass  slope
in denser environments, at least for red galaxies, while others
\citep[e.g.][]{Croton2005,Hoyle2005,McNaught-Roberts2014,Mortlock2015}
find no such dependence.


Turning now to the GAMA mock catalogues, the LFs have a higher abundance
of faint satellite galaxies in massive haloes compared with GAMA data.
It is beyond the scope of this paper to explore the physical reasons
for this, but we do compare predictions of two more recent
SAMs with GAMA data in Riggs et al. (in prep.).
Standard Schechter functions systematically underestimate the
faint-end of the LF in all but \mass1 mock groups, and so the resulting
parameter fits should be treated with due caution.

The SMF shape in all simulations is generally independent of halo mass
at low stellar masses, $\lg \mass_* \la 10.5$, whereas the GAMA
low-mass slope is steeper in \mass1 groups.
Only at high stellar masses, $\lg \mass_* \ga 10.5$,
do the simulations reveal an increasing number density in higher-mass haloes.
Schechter function fits are unable to capture this behaviour, under-fitting
the high-mass end in all simulations, and
showing minimal dependence of $\mass^*$ and $\alpha$ on halo mass.
Varying the rate of the high-mass decline via the $\beta$ parameter in 
equation~(\ref{eqn:schec}) cannot eliminate this discrepancy.

A double Schechter function is required in order to fit a
low-mass upturn in the \lgal\ SMF for $\lg \mass_* \la 9.5$:
single Schechter fits are too shallow at the low-mass end.
None of the \tng\ SMFs show significant evidence of a low-mass upturn,
but are very poorly-fit at the high-mass end, particularly for blue
satellites.
The red satellite SMFs are roughly consistent between \tng\ and GAMA,
whereas blue satellites in \tng\ show a large excess at the high-mass end.
It thus appears that \tng\ under-estimates the quenching of massive satellite
galaxies in group environments.
The EAGLE satellite SMF is consistent with \tng\,
but limited to lower masses, $\lg \mass_* \la 11$.

\subsection{Evolution in group environments}

In order to study the effect of group environment on LF/SMF evolution,
we compare in Fig.~\ref{fig:clf_ev} to the (environment-independent) field,
after renormalizing the field galaxy numbers to the number of grouped galaxies
of particular type and environment.
We find that, with the exception of red galaxies,
faint and low-mass galaxies are relatively less abundant
in group environments at low-redshift, $z < 0.1$.
Conversely, luminous and massive galaxies, mostly seen at redshifts $z > 0.1$,
are relatively more common in group environments.
The dominant evolutionary effect in group environments is an increasing
red fraction with decreasing redshift, relative to the field.
The fact that this is seen at redshifts $z < 0.3$ suggests that environment
quenching of galaxies in groups is an ongoing process.

\subsection{Comparison with previous results}

The overall trend of finding more luminous and massive galaxies in higher-mass
haloes can be understood in the context of the hierarchical model
of galaxy formation.
In this model, massive galaxies accrete much
of their stellar mass from sub-haloes, via major and minor mergers
(e.g. \citealt{White1978,Cole2000}).
Analysing the Illustris simulation, \citet{Rodriguez-Gomez2016}
find that while the fraction of stellar mass contributed by accreted stars
is only about 10 per cent for Milky Way-sized galaxies,
it can be more than 80 per cent for $\mass_* \sim 10^{12} \mass_\odot (h=0.7)$
galaxies.
It does not automatically follow, however, that the richer environments
of massive groups will lead to a higher merger rate,
and hence more massive and luminous galaxies.
In fact, mergers are expected to be less frequent in high-mass haloes due to
the large relative galaxy velocities in these environments
\citep{Ostriker1980,Binney1987}.
However, \citet{Sheen2012} have found that 38 per cent of early-type
galaxies in four massive galaxy clusters show evidence of strong
merger features (tidal tails, shells, etc.), comparable to what is found in
low-density field environments \citep{vanDokkum2005}.
\citet{Oh2018} find that 20 per cent of galaxies observed in rich clusters
show post-merger signatures, whereas only 4 per cent show evidence of
ongoing mergers, in agreement with \citet{Sheen2012}.
\citet{Oh2018} suggest that the mergers took place before galaxy accretion
into the cluster environment, a claim supported by the numerical simulations
of \citet{Yi2013}.

The increasing characteristic luminosity and stellar mass
of galaxies with the mass of their
host dark matter halo thus suggests that mergers have been most common
in the past history of galaxies accreted into massive haloes.
\citet{Tomczak2017}, using a semi-empirical model of SMF evolution,
show that the majority of galaxies in high-density regions at
redshift $z \approx 0.8$ are formed from mergers.
We have attempted to extend their model to $z \approx 0.2$,
appropriate for the GAMA data, but find that the evolved SMF,
even at masses $\lg \mass_* \sim 10$,
is very sensitive to the low-mass cutoff chosen for the initial SMF power law.
We therefore prefer to compare our results with more detailed models and
simulations such as \lgal, EAGLE and \tng, finding broad agreement in the
halo-mass dependence of more massive galaxies.

Galaxies built from multiple mergers are more likely to be spheroidal
in morphology \citep{Rodriguez-Gomez2016}.
Moreover, \citet{Rodriguez-Gomez2017} find that mergers play an important role
in determining galaxy morphology in massive ($\mass_* \ga 10^{11} M_\odot$)
galaxies in the Illustris simulation,
with gas-poor mergers promoting the formation of spheroidal galaxies.
In support of this,
\citet{Man2016} and \citet{Mundy2017} estimate from observations that
about one third of the stellar mass in massive galaxies is acquired via
major mergers since redshift $z \approx 3.5$,
This merger-driven scenario naturally explains the domination of the bright
and high-mass ends of the group LF and SMF by spheroidal galaxies.
One should, however, bear in mind that other mechanisms are also likely
to come into play in the formation of spheroids, such as `inside-out' quenching
\citep{Tacchella2018}.

Unlike \citet{Yang2008,Yang2009}, we find no evidence of a systematic 
steepening with halo mass of faint-end/low-mass LF/SMF;
in fact the lowest-mass \mass1 haloes tend to have the steepest slopes.
We caution that apparent trends of Schechter-like parameters should be
treated with caution, and can be misleading in cases when the
fitting function poorly fits the data.
One should also bear in mind that galaxies have had longer to interact in
low-redshift (mostly low-mass) haloes, and that this is unlikely to be
accounted for by our global evolution corrections.

\subsection{Caveats and future prospects}

The high spectroscopic completeness of the GAMA survey,
and the minimum group membership requirement ($N_{\rm FoF} > 4$), should result
in a higher-fidelity group catalogue than the much larger SDSS catalogue
of \citet{Yang2007}.
However, the GAMA groups are by no means perfect.
In particular, comparison with mocks (Fig.~\ref{fig:mass_comp}) suggests
that low-mass haloes ($\lg \mass_h \la 13.5$) have 
masses overestimated by $\Delta \lg \mass_h \approx 0.5$ dex.
This leads to some small systematic errors in the halo-dependent LF and SMF,
particularly at the faint/low-mass end (Appendix~\ref{sec:fofvhalo}).

Because we analyse a flux-limited sample of GAMA groups,
we cannot separate the effects of host halo mass and (observed)
group membership.
A large, volume-limited sample of galaxy groups would enable more reliable
conclusions on the effects of group environment to be drawn.
Such a sample will be provided by the upcoming Wide Area VISTA Extragalactic
Survey (WAVES; \citealt{Driver2019}).

\section*{Acknowledgements}

GAMA is a joint European-Australasian project based around a
spectroscopic campaign using the Anglo-Australian Telescope. The GAMA
input catalogue is based on data taken from the Sloan Digital Sky
Survey and the UKIRT Infrared Deep Sky Survey. Complementary imaging
of the GAMA regions is being obtained by a number of independent
survey programs including GALEX MIS, VST KiDS, VISTA VIKING, WISE,
Herschel-ATLAS, GMRT and ASKAP providing UV to radio coverage. GAMA is
funded by the STFC (UK), the ARC (Australia), the AAO, and the
participating institutions. The GAMA website is
http://www.gama-survey.org/.

JAVM was supported by the Mexican National Council for Science and
Technology (CONACyT) scholarship scheme, and thanks Peder Norberg
for useful discussions and hospitality.
JL (ORCID 0000-0001-5290-8940) acknowledges support from the
Science and Technology Facilities Council (STFC)
(grant number ST/I000976/1).
SDR is supported by a STFC studentship.

We acknowledge the Virgo Consortium and the \tng\ team for making their
simulation data available.
The EAGLE simulations were performed using the DiRAC-2 facility at Durham,
managed by the ICC, and the PRACE facility Curie based in France at TGCC, CEA,
Bruy\`{e}res-le-Ch\^{a}tel.
This work also used the 2015 public version of the Munich model of
galaxy formation and evolution: \lgal.
The source code and a full description of the model are available at
\url{https://lgalaxiespublicrelease.github.io/}.

Finally, we thank two anonymous referees for their careful reading of the
manuscript and helpful suggestions for improvement.

\section*{Data Availability}

The data underlying this article will be shared on reasonable request
to the corresponding author.
Tabulated LF and SMF results will be made available via the GAMA website
\url{http://www.gama-survey.org/}.

\bibliographystyle{mnras}
\bibliography{library}

\appendix

\section{Comparison of field LFs and SMFs}  \label{sec:field_comp}

We here compare the field (environment-independent) LFs and SMFs for
the GAMA data, the FoF mocks, the \lgal\ SAM,
and two hydrodynamical simulations.

\subsection{GAMA, mock and TNG $r$-band field LFs}

\begin{figure} 
\includegraphics[width=\linewidth]{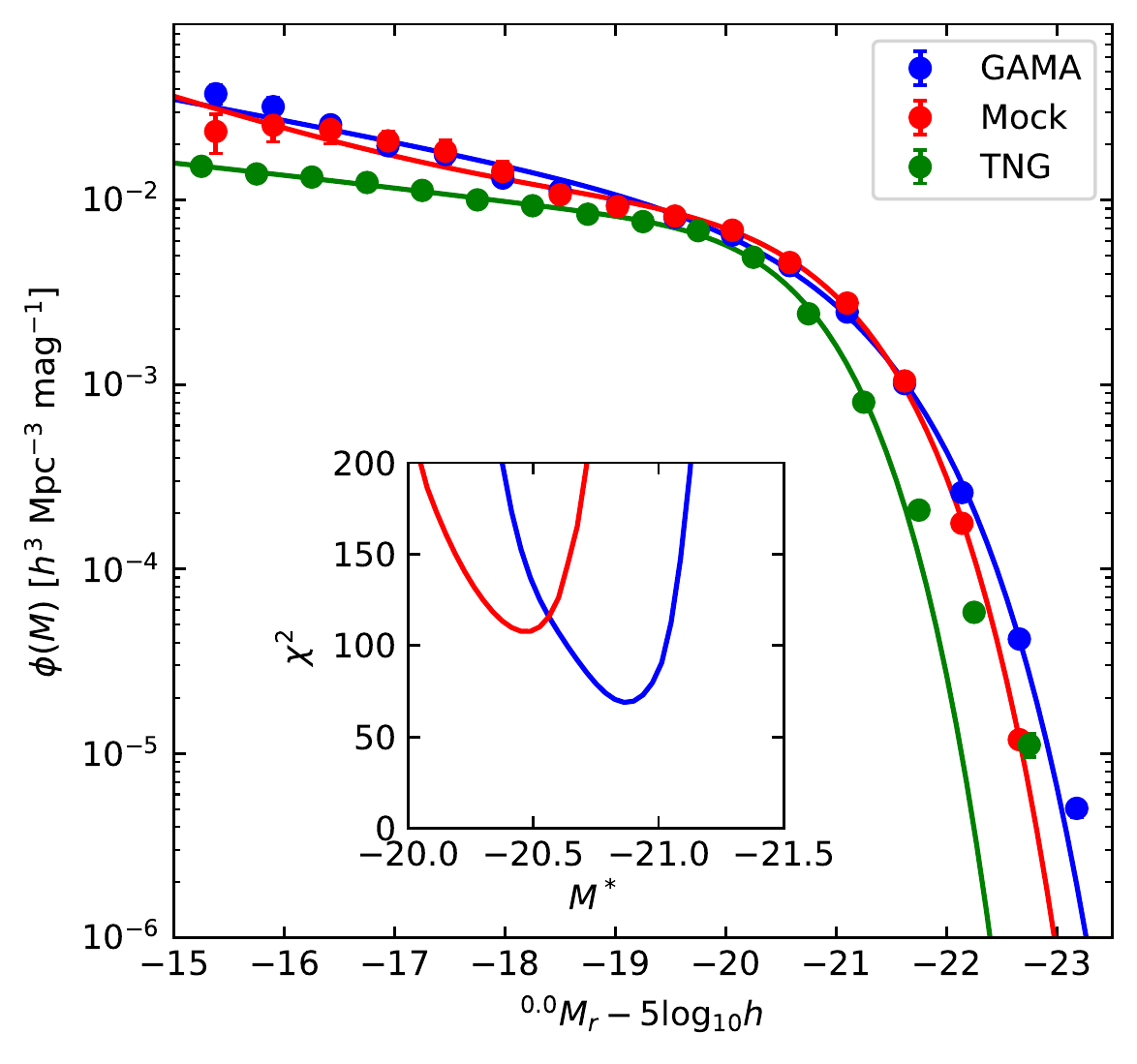}
\caption{Field galaxy $r$-band LFs for GAMA, FoF mock and TNG.
  Symbols show $1/\Vmax$ estimates (density-corrected for GAMA),
  and lines show double Schechter function fits.
  The inset shows GAMA and mock $\chi^2$ profiles for characteristic
  magnitude $M^*$;
  in each case there are 12 degrees of freedom.
  }
\label{fig:lf_field_comp}
\end{figure}

We compare the $r$-band field LFs from GAMA, FoF mocks and \tng\ in
Fig.~\ref{fig:lf_field_comp}.
For GAMA, we use a density-corrected \Vmax\ estimator \citep{Loveday2015}
with errors determined from jackknife sampling.
Mock LFs are estimated using a standard \Vmax\ estimator and errors show
the standard deviation between the nine mocks.
We see that the brightest galaxies in the mocks are not as bright as those
in the GAMA data.
Fitting a double Schechter function (equation 6 of \citealt{Baldry2012})
to the binned LFs, we find that the
characteristic magnitude $M^*$ of GAMA galaxies is about 0.5 mag brighter than
in the mocks.
The GAMA binned LF faint-end slope is also slightly steeper,
although note that the
Millennium-based mock catalogues are not expected to be fully complete fainter
than $M_r \approx -17$ mag.
These differences between the GAMA and mock LFs should be borne in mind
when comparing group LF results.
In particular, one should not focus on differences between GAMA and mocks
in any given halo mass bin, but instead compare the {\em trends}
with halo mass for the real and mock data.

The TNG LF is obtained using the $z=0.2$, TNG300-1 synthetic stellar
photometry catalogue,
which uses dust model C from \citet{Nelson2018}.
We see that while TNG gives a good match to GAMA around the characteristic
magnitude ($M_r - 5 \log h \approx -20$) and at the extreme bright end
($M_r - 5 \log h \approx -23$),
it predicts far too few faint and moderately bright galaxies.
We also note that even a double Schechter function is unable to match the
shape of the \tng\ LF at the bright end.

\subsection{GAMA versus simulated field SMFs}

\begin{figure} 
\includegraphics[width=\linewidth]{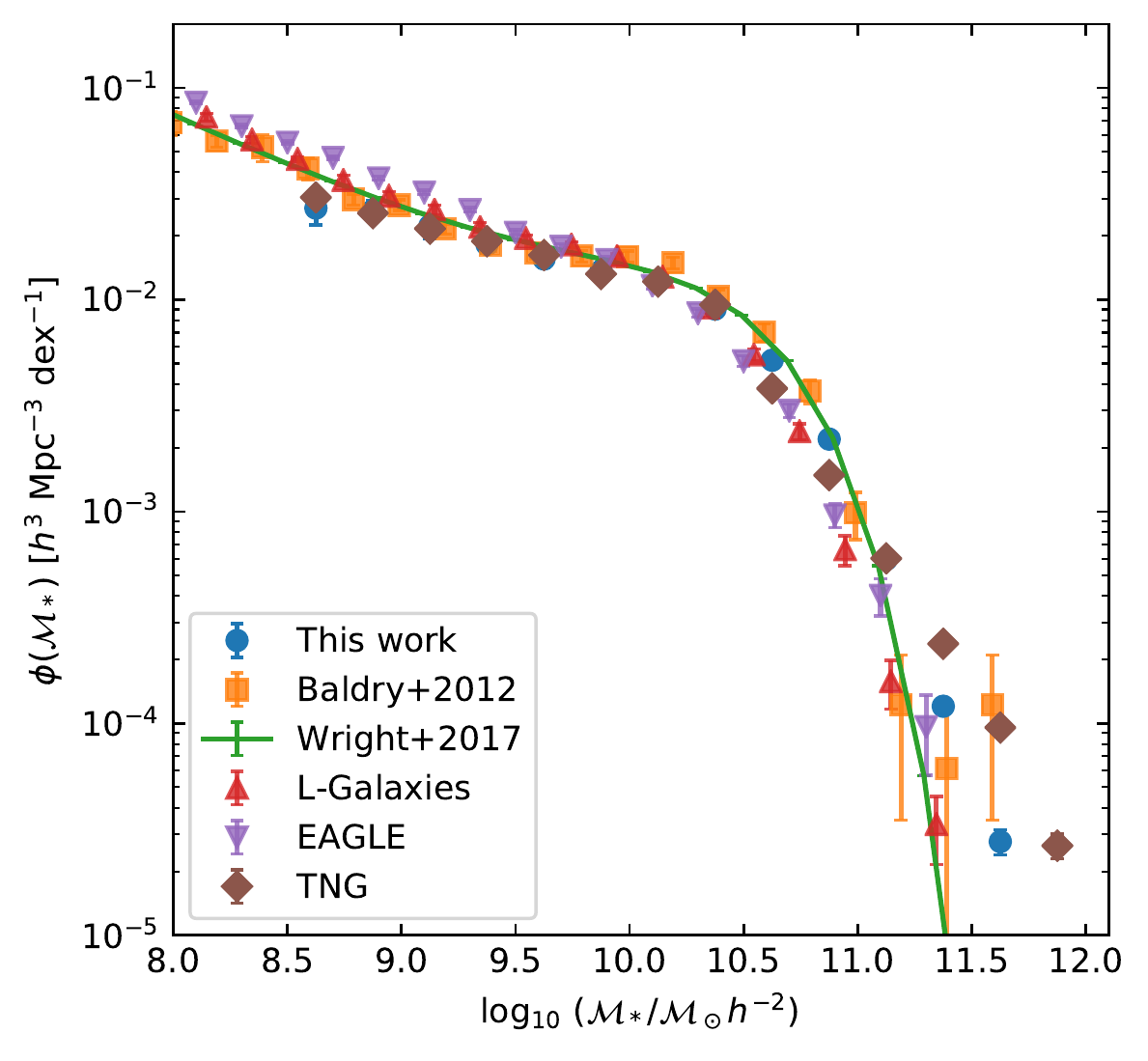}
\caption{Field galaxy SMFs for GAMA estimated by the present work and two
  previous GAMA papers, along with predictions from the \lgal\ SAM
  and the EAGLE and \tng\ hydrodynamical simulations.
}
\label{fig:smf_field_comp}
\end{figure}

We compare the field SMFs from GAMA data, the \lgal\ SAM,
and the EAGLE and \tng\ hydrodynamical simulations in
Fig.~\ref{fig:smf_field_comp}.
We see that our estimate of the GAMA field galaxy SMF agrees well
with previous GAMA estimates by \citet{Baldry2012} and
\citet[their double Schechter function fit]{Wright2017}.
The simulations agree well with the GAMA observations, except that
\tng\ over-predicts the numbers of very massive
($\lg \mass_* \ga 11$) galaxies.
This over-abundance of simulated, very massive galaxies suggests that
the effects of AGN feedback in \tng\ may be underestimated in such hosts.
This would also at least partly explain the \tng\ high-mass excess seen in
group environments (Fig.~\ref{fig:csmf}).

\section{FoF versus halo mock LF results}  \label{sec:fofvhalo}

\begin{figure}
\includegraphics[width=\linewidth]{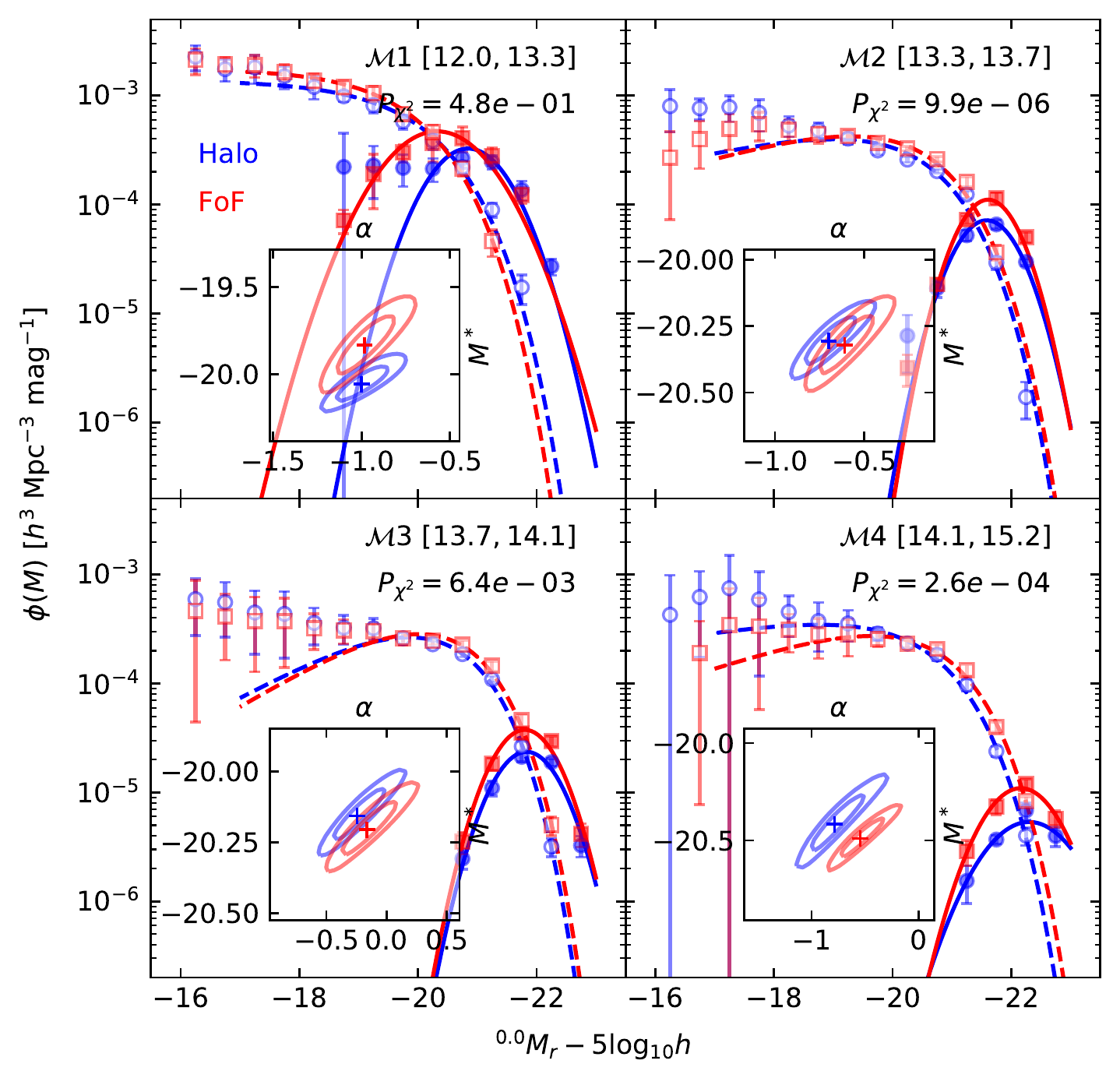}
\caption{Halo mass dependent LFs for halo (blue circles) and FoF (red squares)
  mocks.
  Filled markers and continuous lines show central galaxy LFs with their
  best-fitting log-normal function.
  Open markers and dashed lines show satellite galaxy LFs with their
  best-fitting Schechter function.
  Insets show $1$- and 2-$\sigma$ likelihood contours on the Schechter
  parameters $\alpha$ and $M^*$.
  Also shown in each panel is the $\chi^2$ probability that the composite
  (central plus satellite)
  non-parametric LFs follow the same distribution.
  The difference is statistically significant for all mass bins apart from
  \mass1.
}
\label{fig:clf_mock_comp}
\end{figure}

We compare group LF results obtained using the FoF and halo mocks
in Fig.~\ref{fig:clf_mock_comp}.
Note that in this comparison, the same mock galaxies have been used.
The only difference lies in how the galaxies are assigned to groups,
and how the masses of the groups are determined.

The composite (central plus satellite) halo and FoF non-parametric LF estimates
are significantly different (exceeding 95 per cent confidence)
for all but the lowest mass bin.
We see that the effect of using FoF group-finding and luminosity-based
masses is to change the best-fitting Schechter parameters
for the satellite galaxies by around 1--3$\sigma$.
The central galaxy LFs appear to be consistent in peak luminosity
apart from in FoF \mass1 haloes, where they are fainter.

FoF mocks tend to have slightly less steep faint-end slopes than halo mocks;
all are significantly under-fitting the faint end, particularly in \mass3
groups.
In the lowest-mass haloes, \mass1, FoF mocks have fainter characteristic
magnitudes $M^*$; in all other haloes, there is no significant difference.

Insofar as the mock catalogues are representative of the GAMA data,
we can infer that the GAMA LF Schechter parameters are likely to be biased by
$\sim 1 \sigma$ in intermediate mass bins,
with slightly worse errors in the lowest- and highest-mass haloes.
When comparing GAMA data with mocks,
we use the FoF mocks, under the assumption that they suffer similar biases
to the GAMA groups.

\section{Testing LF and CLF estimators}  \label{sec:simtests}

In this appendix we compare estimates of the LF and CLF from simple simulations
of known CLF, showing that the LF may be recovered without bias,
but that an unbiased CLF estimate is only possible from a volume-limited sample.
We first describe our CLF estimator (our LF estimator is described in
Section~\ref{sec:vmax_est}), and then describe the generation of simple
group and galaxy catalogues with known CLF.
Finally, we present and discuss the recovered CLFs and LFs from
these simulations.

\subsection{CLF estimator}  \label{sec:cond_est}

To estimate CLFs for a given bin of halo mass,
we consider the member galaxies of the groups in that mass bin.
The CLF is given by the absolute magnitude histogram of the member galaxies,
weighting each galaxy by the reciprocal of the number of groups in the
halo mass bin in which the galaxy could in principle be observed.
In other words,
for galaxy $i$ which would be visible to redshift $z_{{\rm lim},i}$,
we count groups that lie in the redshift range
$[z_{\rm lo}, \min(z_{\rm hi},\; z_{{\rm lim},i})]$,
where $(z_{\rm lo}, z_{\rm hi})$ are the sample redshift limits.
This is equivalent to the `direct matching' method of
\citet[][equation 4]{Guo2014c}, except that we normalize the CLF on
a per-galaxy basis, rather than on a per-magnitude bin or mass bin basis,
thus allowing account to be taken of individual galaxy $K$-corrections.

Note that by normalising the galaxy counts by the number of groups
in which each galaxy could be seen, one automatically corrects
for radial density variations,
whether due to large-scale structure or number density evolution,
assuming that group counts vary in the same way as galaxy counts,
and so no explicit corrections for $\Delta(z)$ and $P(z)$ are needed.
One would still apply luminosity evolution corrections for a sample in which
luminosity is evolving.

\subsection{Simulated group and galaxy catalogues}

50,000 group masses 
are chosen at random over the range $\lg \mass_h = [12, 15]$
from a Schechter mass function
with somewhat arbitrary, but not unreasonable, shape parameters
$\alpha_{\cal M} = -1$, $\lg \mass_h^* = 14$.
Each group is assigned a redshift randomly drawn from a distribution
that is uniform in comoving volume over the redshift range $z = [0.002, 0.5]$.
These group masses and redshifts are written out to a simulated group catalogue.

Within each group we generate galaxies with luminosities drawn at random
over the absolute magnitude range $M = [-24, -15]$ mag from
Schechter functions whose parameters vary with group mass $\lg \mass_h$
as follows:
\begin{align}  \label{eqn:sim_clf}
  \alpha &= -1.4 - 0.2 \Delta \mass, \nonumber \\
  M^* &= -21.0 - 0.5 \Delta \mass,\\
  \lg \phi^* &= 1.0 + 0.5 \Delta \mass, \nonumber 
\end{align}
and where $\Delta \mass = \lg \mass_h - \lg \mass_h^*$.
These CLF parameters are chosen to roughly match the satellite CLFs of
\citet{Yang2008}.
The number of galaxies generated in each group is chosen at random from a
Poisson distribution whose mean is given by integrating the group's
CLF over the magnitude range $M = [-24, -15]$ mag.
Galaxies are assigned the same redshift as their host group,
and apparent magnitudes are calculated using the same $K-$corrections
as the GAMA mock catalogues (R11, equation 8), but with no evolution.
We write out a simulated galaxy catalogue containing those galaxies with
apparent magnitude $m < 19.8$ mag.
On average, about 2730 groups in each simulation contain five or more
visible member galaxies, comparable with the number of GAMA groups in our
observed sample.

Altogether, nine simulated group and random catalogues are generated.
While these simulated catalogues do not attempt to model imperfections
in group finding or mass estimation,
they do allow us to investigate any biases in the recovered LFs or CLFs
caused by sample selection effects, particularly those associated with
requiring observed groups to have a minimum galaxy membership.

\begin{figure} 
\includegraphics[width=\linewidth]{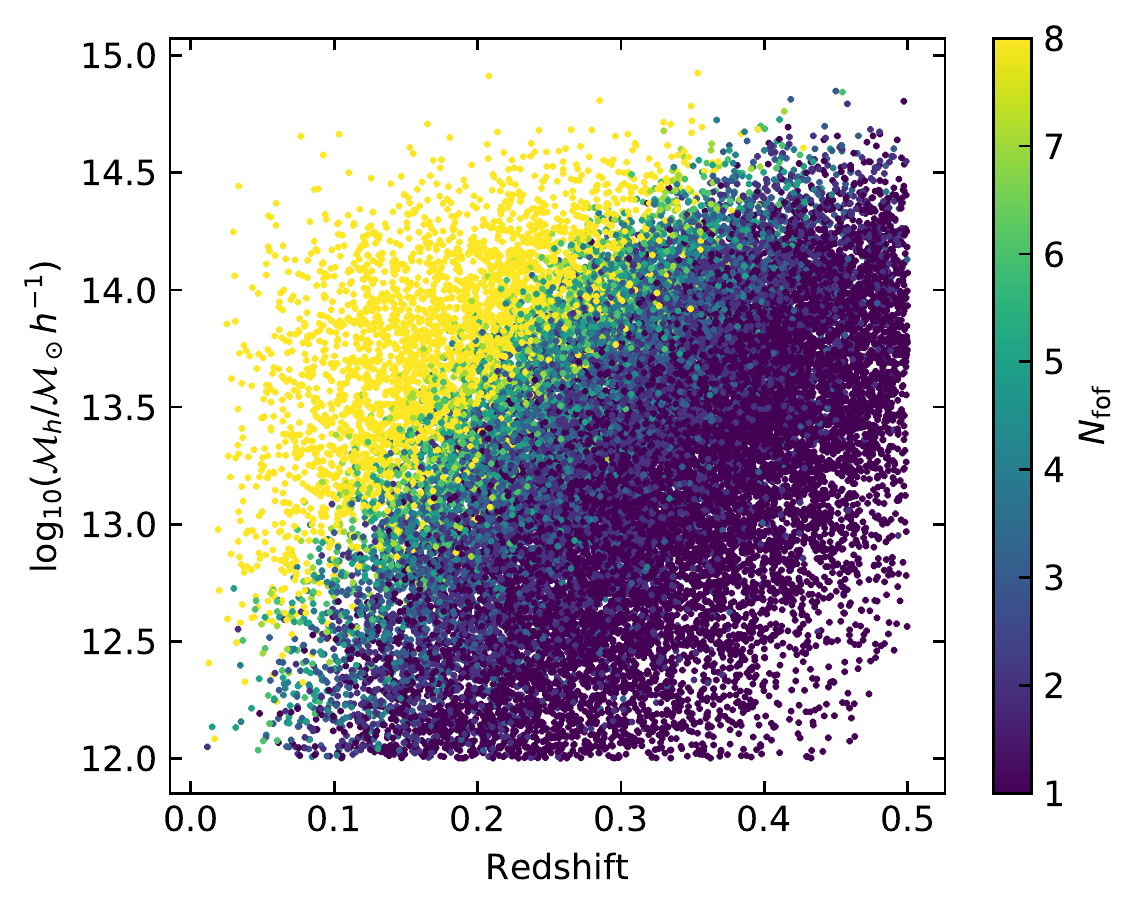} 
\caption{Halo mass--redshift distribution of around 22,000 simulated groups
  chosen at random from those with at least one visible member galaxy.
  Colour-coding indicates the number of visible group members,
  with yellow corresponding to eight or more members.
}
\label{fig:mass_z_sim}
\end{figure}

Fig.~\ref{fig:mass_z_sim} plots the halo mass--redshift distribution for our
simulated groups that contain at least one visible galaxy.
It is clear that low mass groups are incomplete at high redshift,
even when only a single observed ($m < 19.8$ mag) member galaxy is required.
With a membership threshold of 5 galaxies (green and yellow points),
this incompleteness extends to all group masses.
Redshift incompleteness is difficult to quantify in an observed sample,
due to significant scatter in the relation between halo mass and $N$th
brightest galaxy luminosity.
The effects of group redshift incompleteness on the recovered LF
and CLF are explored in the next section.

\subsection{Recovered LFs and CLFs}

\begin{figure} 
\includegraphics[width=\linewidth]{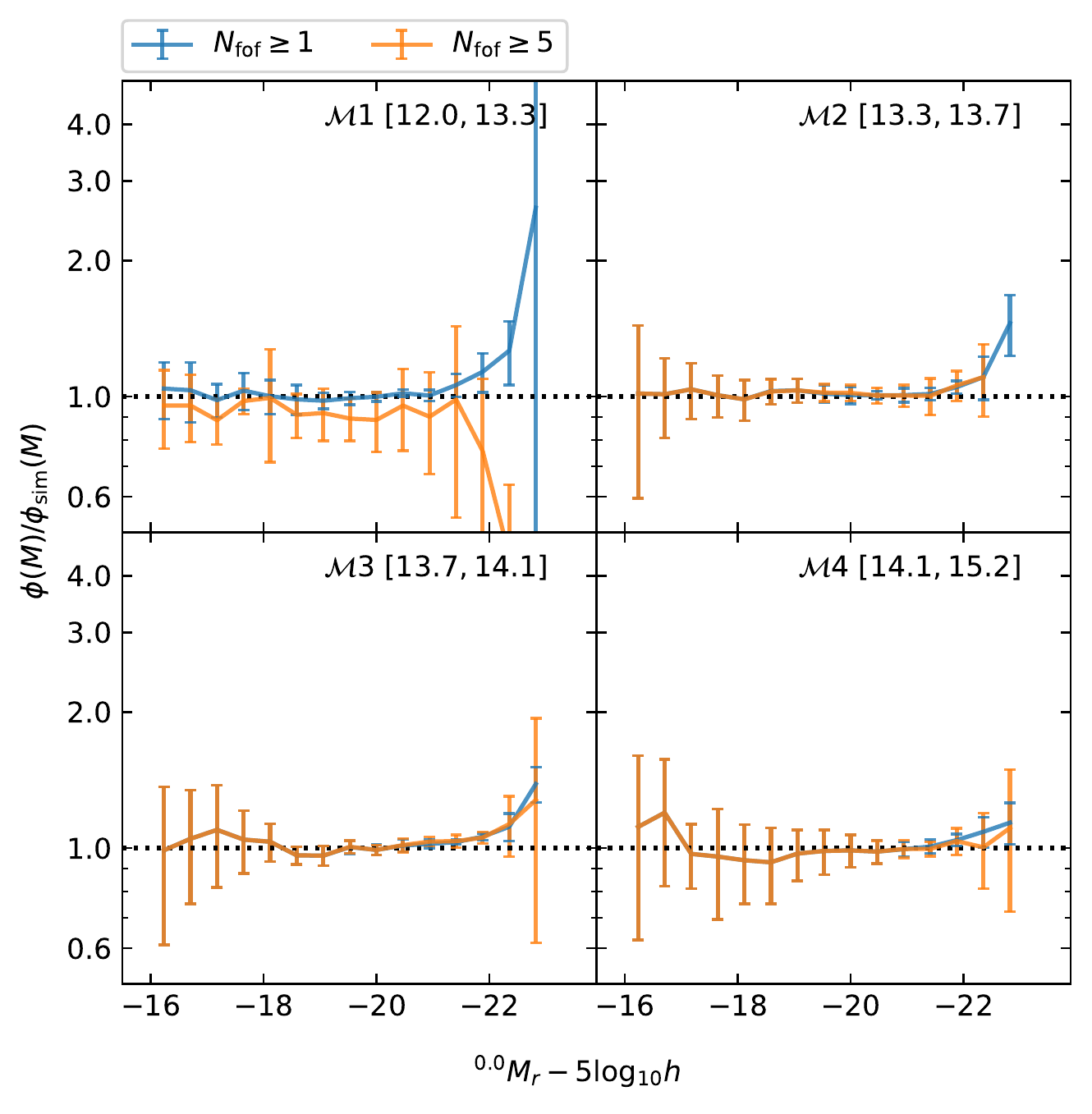} 
\caption{Ratio of recovered to input LF from simulated catalogues.
  Blue and orange error bars, representing the RMS scatter between simulations,
  show LF ratios for groups
  with at least one and five detected members, respectively.
}
\label{fig:clf_sim_vmax}
\end{figure}

\begin{figure} 
\includegraphics[width=\linewidth]{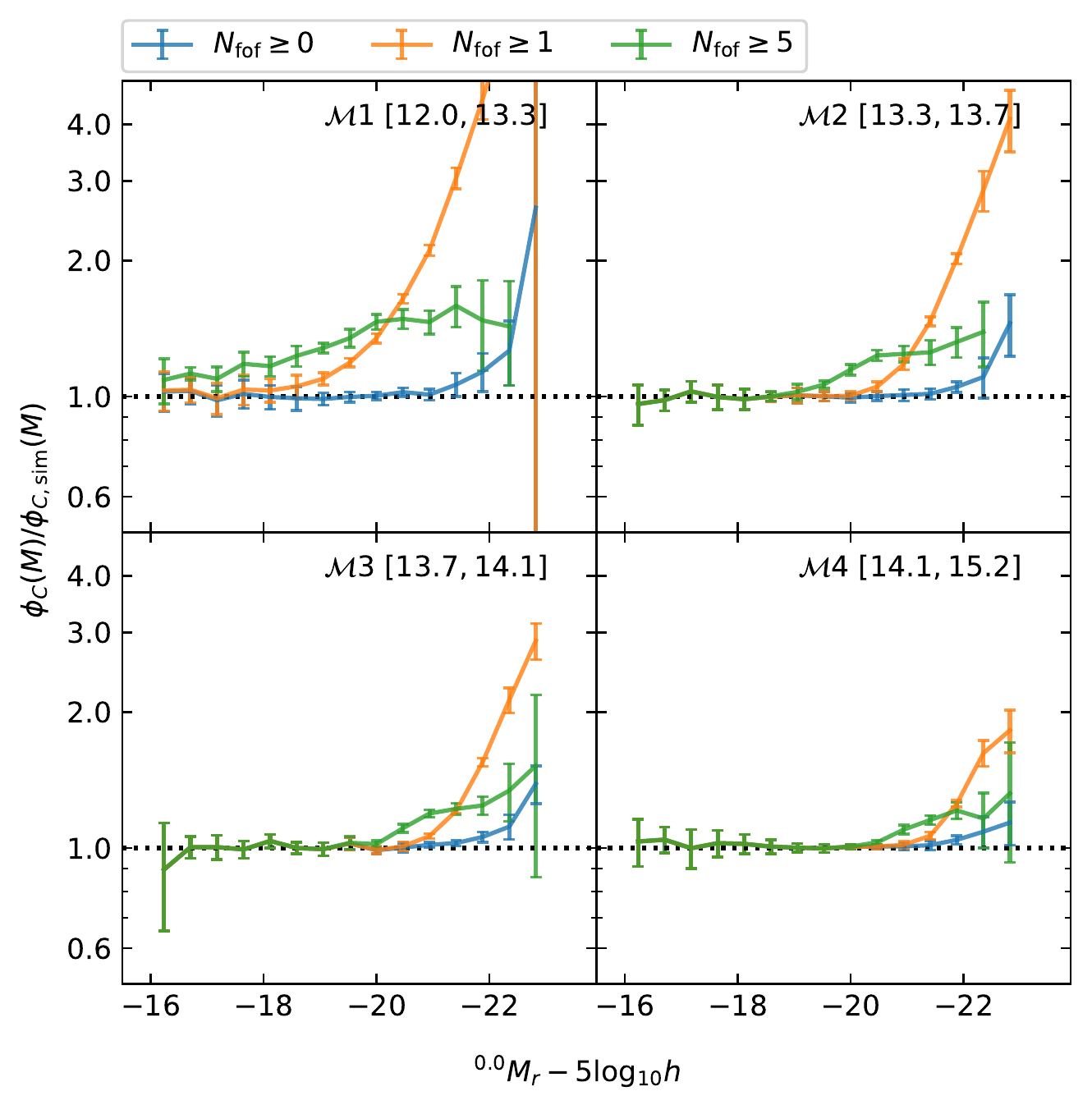} 
\caption{Ratio of recovered to input CLF from simulated catalogues.
  Blue, orange, and green error bars show CLF ratios for groups
  with at least zero, one and five detected members, respectively.
}
\label{fig:clf_sim}
\end{figure}

LFs and CLFs are determined from our simulated catalogues in the same way
as for the GAMA mocks.
The ratios of the recovered to simulated LFs and CLFs, (evaluated
individually for each group, and then summed within mass bins), are shown in 
Figs.~\ref{fig:clf_sim_vmax} and \ref{fig:clf_sim} respectively.
For the LF ratio plots, we have rescaled the input (per group) CLFs to
($1/\Vmax$-weighted) LFs by the factor $N_{\rm group}/V$,
where $N_{\rm group}$ is the {\em total} number of groups simulated
in each mass bin,
and $V \approx 3.6 \times 10^7 \volunit$ is the effective volume
of the simulations, assuming that they cover the same sky area as the
GAMA mocks.
It is not possible to scale observed LFs and CLFs in this way,
since we do not know the total number of groups in each mass bin,
only the number that are observed.

The ($\Vmax$-normalized) LFs (Fig.~\ref{fig:clf_sim_vmax})
are recovered with near-zero bias, albeit with large scatter in low-mass groups,
even when a minimum group membership of five galaxies is imposed.
By lowering the group membership threshold, we are able to constrain the
LF to {\em brighter} magnitudes, with little improvement to the faint-end
estimates.

The recovered (group-normalized) CLFs (Fig.~\ref{fig:clf_sim})
only do a good job in matching
the simulation input when {\em all} simulated groups (blue lines),
including even those that contain no visible galaxies,
are included in the normalisation of the CLF.
By including only groups with one or more observed members (orange lines),
the bright end of the CLF is overestimated, particularly in low-mass groups.
This overestimation extends to fainter magnitudes as
group membership cut is increased to five or more galaxies,
although interestingly is then less severe at brighter magnitudes.
Overestimation of the CLF occurs because luminous galaxies are visible to
high redshift, at which the group sample is increasingly incomplete with
decreasing mass and increasing membership threshold (Fig.~\ref{fig:mass_z_sim}).
Since there is a wide scatter in the correlation of halo mass
with $N$th brightest galaxy luminosity,
redshift incompleteness is difficult to quantify.
For observed group catalogues, one {\em only} knows of those
groups that have at the very least one member, and so without modelling of the
halo mass function, one cannot calculate a reliable CLF without imposing
stringent redshift limits on the sample.

\begin{figure} 
\includegraphics[width=\linewidth]{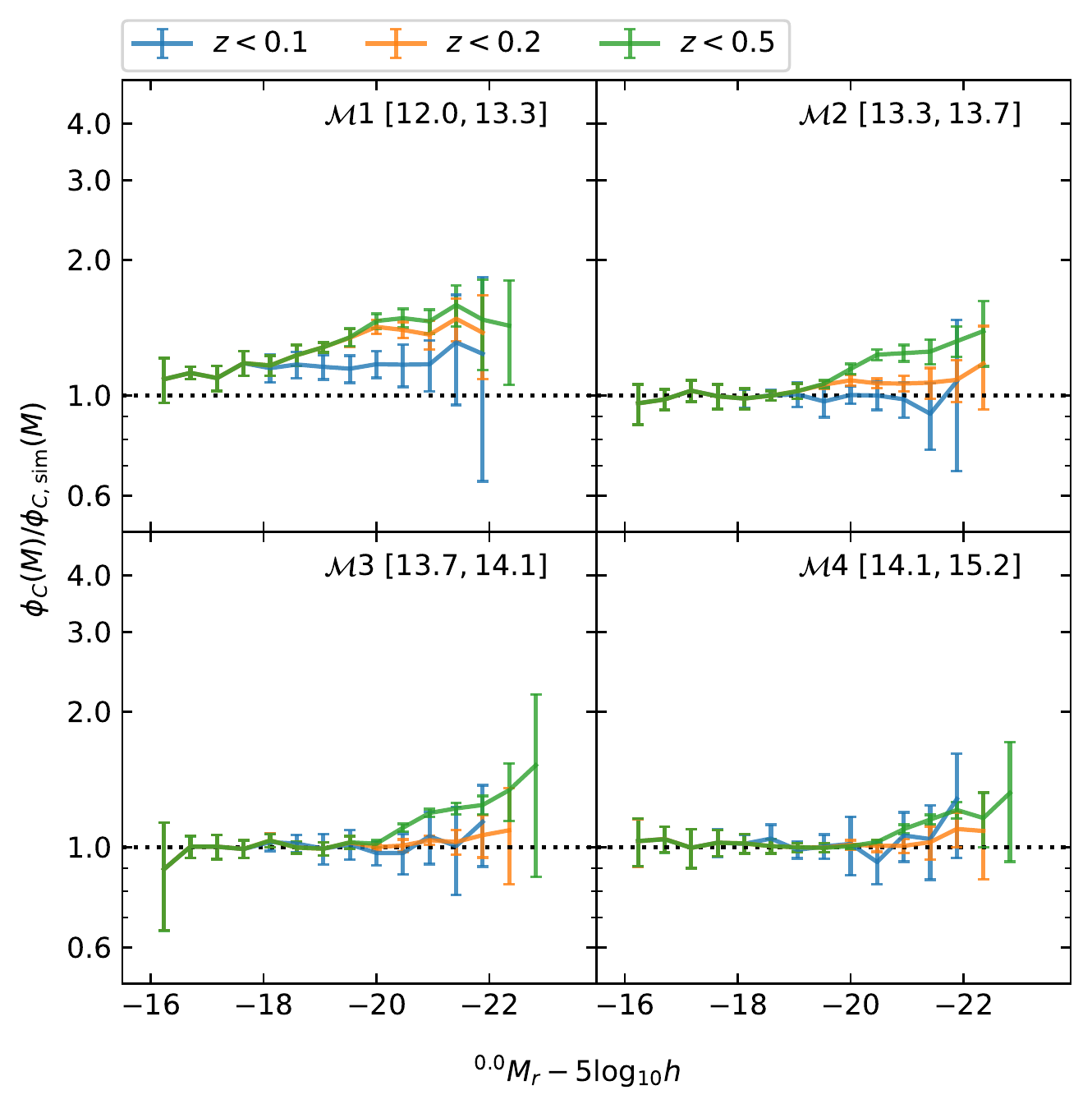} 
\caption{Ratio of recovered to input CLF from simulated catalogues
  for groups with five or more members.
  Blue and orange error bars show CLF ratios after applying
  redshift cuts of 0.1 and 0.2 respectively.
  The green error bars are the same as in Fig.~\ref{fig:clf_sim}.
}
\label{fig:clf_sim_zlim}
\end{figure}

We have re-evaluated the CLFs for groups with five or
more members applying redshift cuts of $z < 0.1$ and $z < 0.2$
(Fig.~\ref{fig:clf_sim_zlim}).
For these simulations, but not necessarily for GAMA groups,
we see that a redshift cut of $z < 0.1$ enables a reliable CLF estimate
for all but the lowest-mass groups.
A less stringent cut of $z < 0.2$ gives acceptable results for mass bins
\mass3 and \mass4.

\begin{figure} 
\includegraphics[width=\linewidth]{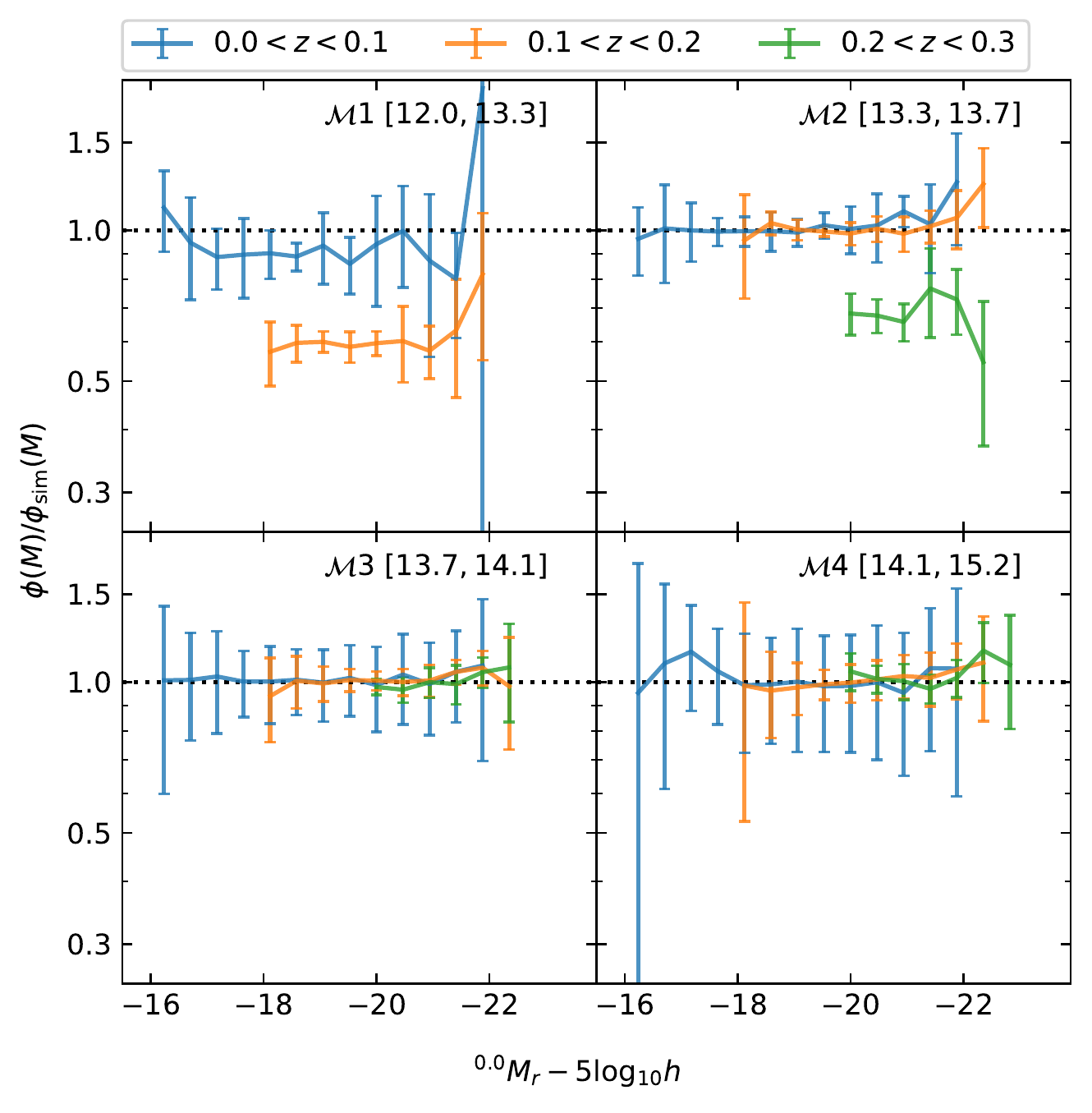} 
\caption{Ratio of recovered to input LF from simulated catalogues
  measured in slices of redshift.
  Blue, orange, and green error bars show LF ratios for groups with
  five or more members at redshift $0.0 < z < 0.1$,
  $0.1 < z < 0.2$, and $0.2 < z < 0.3$, respectively.
}
\label{fig:clf_sim_z}
\end{figure}

Having established that we can recover $\Vmax$-normalized LFs without bias
for all group masses without any redshift cuts,
we investigate the effects of redshift selection on the recovered LF in
Fig.~\ref{fig:clf_sim_z}.
For higher-mass groups, \mass3 and \mass4, the LFs in all three redshift
ranges are recovered without bias.
For mass bin \mass2, the LFs in the highest-redshift range,
$0.2 < z < 0.3$, are biased low.
For the lowest-mass groups, \mass1, the intermediate redshift
range, $0.1 < z < 0.2$, is biased low; there are too few galaxies
at higher redshifts to measure an LF at all.
We conclude that it should be possible to constrain LF evolution in
$\lg \mass_h \ga 13.7$ groups out to redshift $z \approx 0.3$.
For groups in mass bin \mass2, evolution can be reliably
constrained to $z \approx 0.2$.
No determination of LF evolution is possible for groups in the lowest mass bin.

In this appendix, we have demonstrated that the $\Vmax$-normalized LFs
may be recovered from GAMA-like data without bias, even when one is
restricted to groups with five or more members, with a significant
incompleteness in mass-redshift space, as seen in Fig.~\ref{fig:mass_z_sim}.
CLFs may be recovered without bias only when one normalizes
by the total number of groups in the relevant mass and redshift range,
including those with no visible galaxy members (clearly impossible for
an observed, flux-limited sample),
or by applying stringent redshift cuts to obtain a sample that is
volume-limited in group mass
(the approach taken by \citealt{Yang2008,Yang2009}).
For that reason, we present $\Vmax$-normalized LFs and SMFs
rather than group-normalized CLFs and CSMFs in this paper.

\bsp

\label{lastpage}

\end{document}

%% file: authors.tex
\author[J.A.~V\'{a}zquez-Mata et al.]
{{\parbox{\textwidth}{\raggedright J.A.~V\'{a}zquez-Mata,$^{1,2}$
J.~Loveday,$^{1}$\thanks{E-mail:~J.Loveday@sussex.ac.uk}
S.D.~Riggs,$^{1}$
I.K.~Baldry,$^{3}$
L.J.M.~Davies,$^{4}$
A.S.G.~Robotham,$^{4}$
B.W.~Holwerda,$^{5}$
M.J.I.~Brown,$^{6}$
M.~E.~Cluver,$^{7,8}$
L.~Wang,$^{9,10}$
M.~Alpaslan,$^{11}$
J.~Bland-Hawthorn,$^{12}$
S.~Brough,$^{13}$
S.P.~Driver,$^{4,14}$
A.M.~Hopkins,$^{15}$
E.N.~Taylor,$^{7}$
A.H.~Wright$^{16,17}$
}}\\
\vspace{0.4cm}\\
{\parbox{\textwidth}{\raggedright 
$^{1}$Astronomy Centre, University of Sussex, Falmer, Brighton BN1 9QH, UK
\\
$^{2}$Institute of Astronomy, National Autonomous University of Mexico,
    Mexico 04510
\\
$^{3}$Astrophysics Research Institute, Liverpool John Moores University, IC2, Liverpool Science Park, 146 Brownlow Hill, Liverpool, L3 5RF, UK
\\
$^{4}$International Centre for Radio Astronomy Research (ICRAR),
    The University of Western Australia, 35 Stirling Highway, Crawley,
    WA6009, Australia
\\
$^{5}$Department of Physics and Astronomy, University of Louisville, Louisville, KY 40292, USA
\\
$^{6}$School of Physics, Monash University, Clayton, Victoria 3800, Australia
\\
$^{7}$Centre for Astrophysics \& Supercomputing, Swinburne University of Technology, Hawthorn, VIC 3122, Australia
\\
$^{8}$Department of Physics and Astronomy, University of the Western Cape, Robert Sobukwe Road, Bellville, 7535, South Africa
\\
$^{9}$SRON Netherlands Institute for Space Research, Landleven 12, 9747 AD, Groningen, The Netherlands
\\
$^{10}$Kapteyn Astronomical Institute, University of Groningen, Postbus 800, 9700 AV, Groningen, The Netherlands
\\
$^{11}$Center for Cosmology and Particle Physics, Department of Physics, New York University, 726 Broadway, New York, NY 10003, USA
\\
$^{12}$Sydney Institute for Astronomy, School of Physics, University of
Sydney, NSW 2006, Australia
\\
$^{13}$School of Physics, University of New South Wales, NSW 2052, Australia
\\
$^{14}$School of Physics \& Astronomy, University of St Andrews, North
    Haugh, St Andrews, KY16 9SS, UK
\\
$^{15}$Australian Astronomical Optics, Macquarie University
105 Delhi Rd, North Ryde, NSW 2113, Australia
\\
$^{16}$Argelander-Institut f\"{u}r Astronomie, Universit\"{a}t Bonn, Auf dem H\"{u}gel 71, 53121 Bonn, Germany
\\
$^{17}$Astronomisches Institut, Ruhr-Universit\"{a}t Bochum, Universit\"{a}tsstr. 150, 44801 Bochum, Germany
\\
}}}

%% file: group_mass.tex
\begin{table*}
\caption{Group bin names and log-mass limits, number of groups and galaxies,
mean log-mass, and mean redshift for GAMA-II groups, intrinsic mock haloes,
and FoF mock groups.
Note that each mock realisation has about 20 per cent smaller volume than
the GAMA-II equatorial fields.
\label{tab:group_mass_def}}
\begin{tabular}{ccccccccccccccccc}
\hline
& & \multicolumn{4}{c}{GAMA} & & \multicolumn{4}{c}{Halo Mocks} & &
\multicolumn{4}{c}{FoF Mocks} \\
\cline{3-6} \cline{8-11} \cline{13-16} \\[-2ex]
& $\lg {\cal M}_{h, {\rm limits}}$ & 
$N_{\rm grp}$ & $N_{\rm gal}$ & $\overline{\lg {\cal M}_h}$ & $\overline z$ & &
$N_{\rm grp}$ & $N_{\rm gal}$ & $\overline{\lg {\cal M}_h}$ & $\overline z$ & &
$N_{\rm grp}$ & $N_{\rm gal}$ & $\overline{\lg {\cal M}_h}$ & $\overline z$ \\
\hline
${\cal M}1$ & [12.0, 13.3] & 712 & 4520 & 13.03 & 0.12 & & 441 & 3133 & 12.98 & 0.12 & & 584 & 3914 & 12.97 & 0.12\\
${\cal M}2$ & [13.3, 13.7] & 856 & 6817 & 13.50 & 0.19 & & 594 & 4971 & 13.51 & 0.19 & & 744 & 5676 & 13.51 & 0.19\\
${\cal M}3$ & [13.7, 14.1] & 722 & 6944 & 13.88 & 0.26 & & 567 & 6146 & 13.88 & 0.25 & & 668 & 6705 & 13.89 & 0.26\\
${\cal M}4$ & [14.1, 15.2] & 422 & 6762 & 14.37 & 0.32 & & 310 & 7688 & 14.34 & 0.29 & & 353 & 6868 & 14.34 & 0.30\\

\hline
\end{tabular}
\end{table*}

%% file: sim_group_mass.tex
\begin{table*}
\caption{Halo samples for \lgal, EAGLE and \tng\ simulations.
The log-mass limits (second column) are chosen to give
mean log masses close to those of GAMA galaxies in corresponding halo mass bins
(see Table~\ref{tab:group_mass_def}).
For each simulation, we give the number of haloes and galaxies, mean log-mass,
and snapshot redshift.
The number of galaxies quoted for \lgal\ comprises only those from Millennium,
not Millennium II, i.e. those with $\lg \mass_* > 9.5$.
EAGLE and \tng\ samples give the number of galaxies with $\lg \mass_* > 8.5$.
\label{tab:sim_group_mass_def}}
\begin{tabular}{cccccccccccccccccccc}
\hline
& & \multicolumn{4}{c}{\lgal} & &
\multicolumn{4}{c}{EAGLE} & & \multicolumn{4}{c}{\tng} \\
\cline{3-6} \cline{8-11} \cline{13-16} \\[-2ex]
& $\lg {\cal M}_{h, {\rm limits}}$ & 
$N_{\rm halo}$ & $N_{\rm gal}$ & $\overline{\lg {\cal M}_h}$ & $z$ & &
$N_{\rm halo}$ & $N_{\rm gal}$ & $\overline{\lg {\cal M}_h}$ & $z$ & &
$N_{\rm halo}$ & $N_{\rm gal}$ & $\overline{\lg {\cal M}_h}$ & $z$ \\
\hline
${\cal M}1$ & [12.8, 13.3] & 44665 & 201538 & 13.00 &  0.11 & & 155 & 1612 & 13.02 & 0.18 & & 3713 & 30570 & 13.05 &  0.20 \\
${\cal M}2$ & [13.3, 13.7] & 11906 & 134384 & 13.47 &  0.18 & & 42 & 1069 & 13.47 & 0.18 & & 1040 & 21909 & 13.50 &  0.20 \\
${\cal M}3$ & [13.7, 14.1] & 3665 & 93949 & 13.86 &  0.26 & & 9 & 644 & 13.85 & 0.18 & & 405 & 19055 & 13.89 &  0.20 \\
${\cal M}4$ & [14.1, 14.8] & 910 & 60985 & 14.29 &  0.31 & & 5 & 950 & 14.31 & 0.18 & & 120 & 15468 & 14.37 &  0.20 \\

\hline
\end{tabular}
\end{table*}

%% file: clf_cen.tbl
\begin{math}
             \begin{array}{crcccc}
             \hline
             & N_{\rm gal} & M_c & \sigma_c & \lg \phi^*_c & \chi^2/\nu \\[0pt]

                  \hline
                  \multicolumn{6}{c}{\mbox{Mock all}} \\
\mass1 & 584 & -20.33\pm 0.06 &  0.75\pm 0.03 & -3.33\pm 0.05 & 11.7/ 6 \\[0pt]
\mass2 & 744 & -21.61\pm 0.02 &  0.45\pm 0.01 & -3.96\pm 0.03 & 16.0/ 8 \\[0pt]
\mass3 & 668 & -21.80\pm 0.03 &  0.47\pm 0.02 & -4.41\pm 0.03 &  9.2/ 6 \\[0pt]
\mass4 & 352 & -22.10\pm 0.04 &  0.47\pm 0.03 & -4.93\pm 0.04 &  7.7/ 6 \\[0pt]

                  \hline
                  \multicolumn{6}{c}{\mbox{GAMA all}} \\
\mass1 & 699 & -20.71\pm 0.08 &  0.55\pm 0.04 & -3.01\pm 0.06 &  3.9/ 5 \\[0pt]
\mass2 & 842 & -21.70\pm 0.02 &  0.43\pm 0.01 & -3.95\pm 0.03 &  2.5/ 3 \\[0pt]
\mass3 & 699 & -22.03\pm 0.02 &  0.41\pm 0.01 & -4.39\pm 0.03 &  8.6/ 3 \\[0pt]
\mass4 & 392 & -22.35\pm 0.02 &  0.47\pm 0.02 & -4.97\pm 0.03 &  3.5/ 2 \\[0pt]

                  \hline
                  \multicolumn{6}{c}{\mbox{GAMA blue}} \\
\mass1 & 215 & -20.45\pm 0.16 &  0.64\pm 0.09 & -3.39\pm 0.11 &  1.5/ 5 \\[0pt]
\mass2 & 230 & -21.51\pm 0.03 &  0.42\pm 0.02 & -4.52\pm 0.04 &  2.6/ 3 \\[0pt]
\mass3 & 189 & -21.95\pm 0.04 &  0.46\pm 0.03 & -5.03\pm 0.06 &  0.6/ 3 \\[0pt]
\mass4 & 101 & -22.30\pm 0.06 &  0.53\pm 0.04 & -5.64\pm 0.05 &  0.6/ 2 \\[0pt]

                  \hline
                  \multicolumn{6}{c}{\mbox{GAMA red}} \\
\mass1 & 484 & -20.80\pm 0.06 &  0.51\pm 0.03 & -3.24\pm 0.06 &  3.2/ 4 \\[0pt]
\mass2 & 612 & -21.77\pm 0.02 &  0.41\pm 0.02 & -4.06\pm 0.04 &  3.1/ 3 \\[0pt]
\mass3 & 510 & -22.09\pm 0.02 &  0.41\pm 0.02 & -4.53\pm 0.03 &  6.8/ 2 \\[0pt]
\mass4 & 291 & -22.37\pm 0.02 &  0.44\pm 0.02 & -5.06\pm 0.03 &  5.7/ 2 \\[0pt]

                  \hline
                  \multicolumn{6}{c}{\mbox{GAMA low-n}} \\
\mass1 & 106 & -20.59\pm 0.10 &  0.41\pm 0.05 & -3.51\pm 0.17 &  2.1/ 3 \\[0pt]
\mass2 & 95 & -21.44\pm 0.06 &  0.50\pm 0.05 & -5.00\pm 0.08 &  0.2/ 3 \\[0pt]
\mass3 & 67 & -21.69\pm 0.05 &  0.41\pm 0.03 & -5.42\pm 0.03 &  0.1/ 2 \\[0pt]
\mass4 & 32 & -21.92\pm 0.12 &  0.58\pm 0.13 & -6.14\pm 0.11 &  0.3/ 2 \\[0pt]

                  \hline
                  \multicolumn{6}{c}{\mbox{GAMA high-n}} \\
\mass1 & 593 & -20.75\pm 0.09 &  0.55\pm 0.04 & -3.12\pm 0.06 &  3.5/ 5 \\[0pt]
\mass2 & 747 & -21.74\pm 0.02 &  0.41\pm 0.01 & -3.97\pm 0.03 &  0.7/ 3 \\[0pt]
\mass3 & 632 & -22.08\pm 0.02 &  0.39\pm 0.01 & -4.41\pm 0.03 &  4.9/ 3 \\[0pt]
\mass4 & 360 & -22.38\pm 0.02 &  0.45\pm 0.02 & -4.99\pm 0.02 &  2.2/ 2 \\[0pt]

        \hline
        \end{array}
        \end{math}

%% file: clf_sat.tbl
\begin{math}
             \begin{array}{crcccc}
             \hline
             & N_{\rm gal} & M^* & \alpha & \lg \phi^*_s & \chi^2/\nu \\[0pt]

                  \hline
                  \multicolumn{6}{c}{\mbox{Mock all}} \\
\mass1 & 3273 & -19.87\pm 0.09 & -1.04\pm 0.07 & -2.72\pm 0.06 &  3.3/ 9 \\[0pt]
\mass2 & 4931 & -20.34\pm 0.07 & -0.65\pm 0.09 & -3.02\pm 0.03 & 20.5/10 \\[0pt]
\mass3 & 6037 & -20.22\pm 0.07 & -0.19\pm 0.16 & -3.08\pm 0.02 & 21.4/11 \\[0pt]
\mass4 & 6515 & -20.44\pm 0.06 & -0.44\pm 0.11 & -3.17\pm 0.03 &  7.3/13 \\[0pt]

                  \hline
                  \multicolumn{6}{c}{\mbox{GAMA all}} \\
\mass1 & 3579 & -19.98\pm 0.13 & -1.02\pm 0.11 & -2.68\pm 0.10 &  9.7/ 8 \\[0pt]
\mass2 & 5757 & -20.32\pm 0.08 & -0.73\pm 0.10 & -3.01\pm 0.04 & 11.2/ 9 \\[0pt]
\mass3 & 6014 & -20.36\pm 0.05 & -0.38\pm 0.08 & -3.15\pm 0.02 & 18.9/10 \\[0pt]
\mass4 & 6108 & -20.83\pm 0.05 & -0.68\pm 0.08 & -3.38\pm 0.02 &  6.9/11 \\[0pt]

                  \hline
                  \multicolumn{6}{c}{\mbox{GAMA blue}} \\
\mass1 & 2260 & -20.28\pm 0.20 & -1.30\pm 0.09 & -3.09\pm 0.11 &  5.9/ 8 \\[0pt]
\mass2 & 2837 & -20.36\pm 0.10 & -0.94\pm 0.10 & -3.37\pm 0.05 & 10.2/ 9 \\[0pt]
\mass3 & 2541 & -20.35\pm 0.08 & -0.56\pm 0.10 & -3.55\pm 0.03 & 12.5/10 \\[0pt]
\mass4 & 2271 & -20.81\pm 0.08 & -0.79\pm 0.12 & -3.85\pm 0.04 &  9.9/11 \\[0pt]

                  \hline
                  \multicolumn{6}{c}{\mbox{GAMA red}} \\
\mass1 & 1319 & -19.69\pm 0.16 & -0.49\pm 0.18 & -3.06\pm 0.10 & 15.7/ 8 \\[0pt]
\mass2 & 2920 & -20.30\pm 0.08 & -0.52\pm 0.08 & -3.27\pm 0.03 &  6.8/ 9 \\[0pt]
\mass3 & 3473 & -20.47\pm 0.06 & -0.43\pm 0.08 & -3.40\pm 0.02 & 16.5/ 9 \\[0pt]
\mass4 & 3837 & -20.81\pm 0.04 & -0.54\pm 0.07 & -3.57\pm 0.02 &  8.9/11 \\[0pt]

                  \hline
                  \multicolumn{6}{c}{\mbox{GAMA low-n}} \\
\mass1 & 2064 & -19.47\pm 0.10 & -1.02\pm 0.11 & -2.79\pm 0.08 &  8.7/ 8 \\[0pt]
\mass2 & 2551 & -20.07\pm 0.08 & -0.97\pm 0.09 & -3.33\pm 0.05 &  4.7/ 9 \\[0pt]
\mass3 & 2188 & -20.06\pm 0.05 & -0.54\pm 0.09 & -3.51\pm 0.02 & 19.6/10 \\[0pt]
\mass4 & 1928 & -20.45\pm 0.06 & -0.79\pm 0.11 & -3.77\pm 0.03 & 11.7/10 \\[0pt]

                  \hline
                  \multicolumn{6}{c}{\mbox{GAMA high-n}} \\
\mass1 & 1515 & -20.22\pm 0.25 & -0.74\pm 0.19 & -3.10\pm 0.13 & 15.6/ 8 \\[0pt]
\mass2 & 3206 & -20.19\pm 0.09 & -0.23\pm 0.12 & -3.20\pm 0.02 &  9.9/ 9 \\[0pt]
\mass3 & 3826 & -20.42\pm 0.06 & -0.25\pm 0.10 & -3.35\pm 0.02 & 14.8/ 9 \\[0pt]
\mass4 & 4180 & -20.81\pm 0.05 & -0.43\pm 0.08 & -3.53\pm 0.02 &  6.9/11 \\[0pt]

        \hline
        \end{array}
        \end{math}

%% file: csmf_comp_cen.tbl
\begin{math}
             \begin{array}{crcccc}
             \hline
             & N_{\rm gal} & \lg\ {\cal M}_c & \sigma_c & \lg \phi^*_c & \chi^2/\nu \\[0pt]

                  \hline
                  \multicolumn{6}{c}{\mbox{GAMA all}} \\
\mass1 & 684 & 10.56\pm 0.03 &  0.30\pm 0.02 & -3.54\pm 0.04 & 28.5/ 7 \\[0pt]
\mass2 & 811 & 10.95\pm 0.01 &  0.25\pm 0.01 & -3.84\pm 0.04 & 13.2/ 5 \\[0pt]
\mass3 & 644 & 11.12\pm 0.01 &  0.22\pm 0.01 & -4.20\pm 0.03 &  3.5/ 3 \\[0pt]
\mass4 & 369 & 11.19\pm 0.01 &  0.24\pm 0.01 & -4.67\pm 0.03 &  9.5/ 3 \\[0pt]

                  \hline
                  \multicolumn{6}{c}{\mbox{GAMA blue}} \\
\mass1 & 200 & 10.35\pm 0.06 &  0.31\pm 0.03 & -4.05\pm 0.05 & 16.1/ 7 \\[0pt]
\mass2 & 202 & 10.72\pm 0.02 &  0.24\pm 0.02 & -4.40\pm 0.06 &  2.6/ 4 \\[0pt]
\mass3 & 143 & 11.00\pm 0.03 &  0.23\pm 0.02 & -4.91\pm 0.05 &  0.7/ 3 \\[0pt]
\mass4 & 80 & 11.12\pm 0.03 &  0.22\pm 0.02 & -5.27\pm 0.08 &  6.1/ 3 \\[0pt]

                  \hline
                  \multicolumn{6}{c}{\mbox{GAMA red}} \\
\mass1 & 484 & 10.67\pm 0.03 &  0.26\pm 0.01 & -3.61\pm 0.05 & 17.0/ 4 \\[0pt]
\mass2 & 609 & 11.02\pm 0.01 &  0.22\pm 0.01 & -3.92\pm 0.04 & 11.9/ 4 \\[0pt]
\mass3 & 501 & 11.15\pm 0.01 &  0.20\pm 0.01 & -4.27\pm 0.03 &  0.9/ 2 \\[0pt]
\mass4 & 289 & 11.24\pm 0.02 &  0.24\pm 0.02 & -4.81\pm 0.04 &  3.3/ 2 \\[0pt]

                  \hline
                  \multicolumn{6}{c}{\mbox{GAMA low-n}} \\
\mass1 & 97 &  9.88\pm 0.25 &  0.64\pm 0.18 & -4.44\pm 0.06 &  1.9/ 4 \\[0pt]
\mass2 & 73 & 10.65\pm 0.05 &  0.25\pm 0.04 & -4.83\pm 0.09 &  3.3/ 2 \\[0pt]
\mass3 & 41 & 10.82\pm 0.04 &  0.21\pm 0.04 & -5.31\pm 0.07 &  0.2/ 1 \\[0pt]
\mass4 & 20 & 10.95\pm 0.07 &  0.22\pm 0.06 & -5.87\pm 0.14 &  0.2/ 1 \\[0pt]

                  \hline
                  \multicolumn{6}{c}{\mbox{GAMA high-n}} \\
\mass1 & 587 & 10.61\pm 0.03 &  0.28\pm 0.01 & -3.56\pm 0.04 & 20.0/ 6 \\[0pt]
\mass2 & 738 & 10.99\pm 0.01 &  0.24\pm 0.01 & -3.85\pm 0.04 & 10.7/ 5 \\[0pt]
\mass3 & 603 & 11.14\pm 0.01 &  0.21\pm 0.01 & -4.20\pm 0.03 &  0.6/ 2 \\[0pt]
\mass4 & 349 & 11.21\pm 0.02 &  0.23\pm 0.01 & -4.69\pm 0.03 &  9.8/ 3 \\[0pt]

                  \hline
                  \multicolumn{6}{c}{\mbox{TNG all}} \\
\mass1 & 3713 & 10.78\pm 0.00 &  0.20\pm 0.00 & -3.06\pm 0.01 & 20.9/ 3 \\[0pt]
\mass2 & 1039 & 11.12\pm 0.01 &  0.18\pm 0.00 & -3.58\pm 0.02 &  8.0/ 3 \\[0pt]
\mass3 & 405 & 11.42\pm 0.01 &  0.18\pm 0.01 & -3.98\pm 0.03 &  7.1/ 2 \\[0pt]
\mass4 & 119 & 11.75\pm 0.02 &  0.19\pm 0.01 & -4.53\pm 0.05 &  0.0/ 1 \\[0pt]

                  \hline
                  \multicolumn{6}{c}{\mbox{TNG blue}} \\
\mass1 & 671 & 10.75\pm 0.01 &  0.22\pm 0.01 & -3.84\pm 0.02 &  0.6/ 2 \\[0pt]
\mass2 & 355 & 11.10\pm 0.01 &  0.19\pm 0.01 & -4.07\pm 0.03 &  1.0/ 3 \\[0pt]
\mass3 & 225 & 11.41\pm 0.01 &  0.18\pm 0.01 & -4.24\pm 0.04 &  7.0/ 2 \\[0pt]
\mass4 & 91 & 11.73\pm 0.02 &  0.20\pm 0.02 & -4.66\pm 0.06 &  0.0/ 1 \\[0pt]

                  \hline
                  \multicolumn{6}{c}{\mbox{TNG red}} \\
\mass1 & 3042 & 10.79\pm 0.00 &  0.19\pm 0.00 & -3.14\pm 0.01 & 13.6/ 3 \\[0pt]
\mass2 & 684 & 11.13\pm 0.01 &  0.17\pm 0.00 & -3.74\pm 0.02 &  8.4/ 3 \\[0pt]
\mass3 & 180 & 11.43\pm 0.01 &  0.17\pm 0.01 & -4.32\pm 0.04 &  1.2/ 2 \\[0pt]
\mass4 & 28 & 11.78\pm 9.99 &  0.16\pm 9.99 & -5.09\pm 9.99 &  0.0/ 0 \\[0pt]

                  \hline
                  \multicolumn{6}{c}{\mbox{LGAL all}} \\
\mass1 & 44636 & 10.52\pm 0.00 &  0.26\pm 0.00 & -3.22\pm 0.00 & 70.1/ 7 \\[0pt]
\mass2 & 11893 & 10.72\pm 0.01 &  0.26\pm 0.00 & -3.78\pm 0.01 & 20.1/ 8 \\[0pt]
\mass3 & 3660 & 10.87\pm 0.01 &  0.25\pm 0.01 & -4.27\pm 0.02 &  6.7/ 9 \\[0pt]
\mass4 & 910 & 11.07\pm 0.02 &  0.25\pm 0.02 & -4.88\pm 0.04 &  2.2/ 5 \\[0pt]

                  \hline
                  \multicolumn{6}{c}{\mbox{EAGLE all}} \\
\mass1 & 155 & 10.66\pm 0.02 &  0.16\pm 0.02 & -2.93\pm 0.06 &  7.3/ 1 \\[0pt]
\mass2 & 42 & 10.88\pm 0.04 &  0.23\pm 0.04 & -3.44\pm 0.10 &  0.4/ 1 \\[0pt]
\mass3 & 9 & 11.05\pm 9.99 &  0.14\pm 9.99 & -4.06\pm 9.99 &  0.0/ 0 \\[0pt]
\mass4 & 5 & 12.00\pm 9.99 &  0.49\pm 9.99 & -3.93\pm 9.99 &  0.3/ 0 \\[0pt]

        \hline
        \end{array}
        \end{math}

%% file: csmf_comp_sat.tbl
\begin{math}
             \begin{array}{crcccc}
             \hline
             & N_{\rm gal} & \lg\ {\cal M}^* & \alpha & \lg \phi^*_s & \chi^2/\nu \\[0pt]

                  \hline
                  \multicolumn{6}{c}{\mbox{GAMA all}} \\
\mass1 & 1882 & 10.31\pm 0.04 & -1.16\pm 0.09 & -3.17\pm 0.07 &  7.4/ 7 \\[0pt]
\mass2 & 3183 & 10.51\pm 0.04 & -0.98\pm 0.09 & -3.27\pm 0.05 &  5.2/ 9 \\[0pt]
\mass3 & 2996 & 10.61\pm 0.03 & -0.84\pm 0.09 & -3.45\pm 0.05 &  9.0/ 9 \\[0pt]
\mass4 & 3147 & 10.77\pm 0.04 & -0.91\pm 0.11 & -3.68\pm 0.06 & 22.5/ 9 \\[0pt]

                  \hline
                  \multicolumn{6}{c}{\mbox{GAMA blue}} \\
\mass1 & 895 & 10.43\pm 0.11 & -1.57\pm 0.14 & -3.77\pm 0.19 &  1.7/ 5 \\[0pt]
\mass2 & 1125 & 10.46\pm 0.09 & -1.10\pm 0.14 & -3.70\pm 0.11 &  4.5/ 9 \\[0pt]
\mass3 & 836 & 10.49\pm 0.06 & -0.85\pm 0.17 & -3.89\pm 0.08 & 11.3/ 9 \\[0pt]
\mass4 & 736 & 10.83\pm 0.07 & -1.06\pm 0.13 & -4.37\pm 0.10 & 15.7/ 9 \\[0pt]

                  \hline
                  \multicolumn{6}{c}{\mbox{GAMA red}} \\
\mass1 & 987 & 10.28\pm 0.04 & -0.84\pm 0.11 & -3.37\pm 0.06 &  6.7/ 7 \\[0pt]
\mass2 & 2058 & 10.47\pm 0.03 & -0.75\pm 0.08 & -3.39\pm 0.04 &  3.7/ 8 \\[0pt]
\mass3 & 2160 & 10.59\pm 0.03 & -0.69\pm 0.08 & -3.56\pm 0.04 &  5.1/ 8 \\[0pt]
\mass4 & 2411 & 10.71\pm 0.03 & -0.74\pm 0.10 & -3.74\pm 0.05 & 13.5/ 9 \\[0pt]

                  \hline
                  \multicolumn{6}{c}{\mbox{GAMA low-n}} \\
\mass1 & 770 & 10.17\pm 0.07 & -1.56\pm 0.13 & -3.57\pm 0.12 &  1.5/ 5 \\[0pt]
\mass2 & 923 & 10.25\pm 0.06 & -1.14\pm 0.15 & -3.58\pm 0.09 &  1.8/ 6 \\[0pt]
\mass3 & 617 & 10.36\pm 0.05 & -1.04\pm 0.16 & -3.91\pm 0.08 &  4.5/ 6 \\[0pt]
\mass4 & 513 & 10.48\pm 0.05 & -0.98\pm 0.16 & -4.17\pm 0.07 &  6.1/ 7 \\[0pt]

                  \hline
                  \multicolumn{6}{c}{\mbox{GAMA high-n}} \\
\mass1 & 1112 & 10.24\pm 0.04 & -0.63\pm 0.12 & -3.23\pm 0.05 &  3.6/ 7 \\[0pt]
\mass2 & 2260 & 10.46\pm 0.04 & -0.58\pm 0.09 & -3.32\pm 0.04 &  7.1/ 9 \\[0pt]
\mass3 & 2379 & 10.57\pm 0.03 & -0.56\pm 0.09 & -3.51\pm 0.04 &  8.4/ 9 \\[0pt]
\mass4 & 2634 & 10.73\pm 0.04 & -0.67\pm 0.10 & -3.72\pm 0.05 & 24.2/ 9 \\[0pt]

                  \hline
                  \multicolumn{6}{c}{\mbox{TNG all}} \\
\mass1 & 26857 & 10.33\pm 0.01 & -0.95\pm 0.01 & -3.06\pm 0.01 & 274.5/ 6 \\[0pt]
\mass2 & 20870 & 10.49\pm 0.01 & -1.02\pm 0.01 & -3.26\pm 0.02 & 119.9/ 7 \\[0pt]
\mass3 & 18650 & 10.56\pm 0.02 & -1.05\pm 0.02 & -3.36\pm 0.02 & 119.1/ 8 \\[0pt]
\mass4 & 15349 & 10.57\pm 0.02 & -1.05\pm 0.02 & -3.45\pm 0.02 & 137.4/ 8 \\[0pt]

                  \hline
                  \multicolumn{6}{c}{\mbox{TNG blue}} \\
\mass1 & 16161 & 10.11\pm 0.01 & -0.83\pm 0.02 & -3.10\pm 0.01 &  2.6/ 6 \\[0pt]
\mass2 & 8377 & 10.17\pm 0.02 & -0.82\pm 0.03 & -3.38\pm 0.02 & 39.0/ 7 \\[0pt]
\mass3 & 4901 & 10.16\pm 0.03 & -0.76\pm 0.04 & -3.60\pm 0.03 & 69.5/ 8 \\[0pt]
\mass4 & 2789 & 10.20\pm 0.04 & -0.84\pm 0.05 & -3.90\pm 0.04 & 50.5/ 8 \\[0pt]

                  \hline
                  \multicolumn{6}{c}{\mbox{TNG red}} \\
\mass1 & 10696 & 10.46\pm 9.99 & -0.91\pm 9.99 & -3.61\pm 9.99 & 1074.3/ 6 \\[0pt]
\mass2 & 12493 & 10.60\pm 0.02 & -1.04\pm 0.02 & -3.58\pm 0.02 & 308.6/ 7 \\[0pt]
\mass3 & 13749 & 10.64\pm 0.02 & -1.08\pm 0.02 & -3.56\pm 0.02 & 178.1/ 8 \\[0pt]
\mass4 & 12560 & 10.62\pm 0.02 & -1.06\pm 0.02 & -3.56\pm 0.02 & 131.9/ 8 \\[0pt]

                  \hline
                  \multicolumn{6}{c}{\mbox{LGAL all}} \\
\mass1 & 156902 & 10.24\pm 0.01 & -0.72\pm 0.02 & -2.88\pm 0.01 & 33.6/ 9 \\[0pt]
\mass2 & 122491 & 10.32\pm 0.01 & -0.79\pm 0.02 & -3.04\pm 0.01 & 24.2/ 9 \\[0pt]
\mass3 & 90289 & 10.36\pm 0.01 & -0.80\pm 0.02 & -3.20\pm 0.01 & 25.3/10 \\[0pt]
\mass4 & 60075 & 10.42\pm 0.02 & -0.88\pm 0.03 & -3.44\pm 0.02 & 27.5/10 \\[0pt]

                  \hline
                  \multicolumn{6}{c}{\mbox{EAGLE all}} \\
\mass1 & 1457 & 10.37\pm 0.07 & -1.17\pm 0.07 & -3.11\pm 0.08 &  9.9/ 6 \\[0pt]
\mass2 & 1027 & 10.54\pm 0.10 & -1.33\pm 0.07 & -3.48\pm 0.12 &  3.6/ 7 \\[0pt]
\mass3 & 635 & 10.22\pm 0.09 & -0.87\pm 0.11 & -3.20\pm 0.09 &  4.9/ 8 \\[0pt]
\mass4 & 945 & 10.43\pm 0.09 & -1.04\pm 0.09 & -3.21\pm 0.09 & 13.7/ 8 \\[0pt]

        \hline
        \end{array}
        \end{math}